\documentclass[preprint2]{aastex6}
\usepackage{multirow}
\usepackage{float}

\usepackage{graphicx}
\usepackage{savesym}
\savesymbol{tablenum}
\usepackage{longtable}
\usepackage{siunitx}
\usepackage{epstopdf}
\graphicspath{ {/Users/emilymoser/Documents/PyCharm/LaTex/Paper_Images/} }

\begin{document}

\title{The High-Mass Protostellar Population of a Massive Infrared Dark Cloud}

\author{Emily Moser\altaffilmark{1},  Mengyao Liu\altaffilmark{2}, Jonathan C. Tan\altaffilmark{2,3}, Wanggi Lim\altaffilmark{4}, Yichen Zhang\altaffilmark{5}, Juan Pablo Farias\altaffilmark{3}}

\altaffiltext{1}{Department of Astronomy, Cornell University, Ithaca, NY, USA}
\altaffiltext{2}{Department of Astronomy, University of Virginia, Charlottesville, VA 22904, USA}
\altaffiltext{3}{Department of Space, Earth and Environment, Chalmers University of Technology, Gothenburg, Sweden}
\altaffiltext{4}{SOFIA Science Center, Mountain View, CA, USA}
\altaffiltext{5}{RIKEN, Tokyo, Japan}

\begin{abstract}
We conduct a census of the high-mass protostellar population of the
$\sim70,000\:M_\odot$ Infrared Dark Cloud (IRDC) G028.37+00.07,
identifying 35 sources based on their $70\:\mu$m emission, as reported
in the {\it Herschel} Hi-GAL catalog of Molinari et al. (2016). We
perform aperture photometry to construct spectral energy distributions
(SEDs), which are then fit with the massive protostar models of Zhang
\& Tan (2018).
We find that the sources span a range of isotropic
luminosities from $\sim$20 to 4,500$\:L_\odot$. The most luminous
sources are predicted to have current protostellar masses of
$m_{*}\sim10\:M_\odot$ forming from cores of mass
$M_{c}\sim40$ to $400\:M_\odot$. The least luminous sources in our sample are
predicted to be protostars with masses as low as $\sim 0.5\:M_\odot$
forming from cores with $M_{c}\sim10\:M_\odot$, which are the minimum
values explored in the protostellar model grid. The detected
protostellar population has a total estimated protostellar mass of
$M_{*}\sim 100\:M_\odot$.
Allowing for completeness corrections, which are constrained by
comparison with an {\it ALMA} study in part of the cloud, we estimate a
star formation efficiency per free-fall time of $\sim3\%$ in the IRDC.
Finally, analyzing the spatial distribution of the sources, we find
relatively low degrees of central concentration of the protostars. The
protostars, including the most massive ones, do not appear to be
especially centrally concentrated in the protocluster as defined by
the IRDC boundary.
\end{abstract}
\section{Introduction}\label{sec:intro}

Interest in the Infrared Dark Clouds (IRDCs) of the Galaxy has grown
dramatically in recent years, as they may inform us about the earliest
stages of massive star and star cluster formation. IRDCs are cold,
dense structures seen against the bright IR emission of the Galactic
plane, with temperatures $T\lesssim25\:$K and H-nuclei number
densities ranging from $n_{\rm H}\sim10^3\:{\rm cm}^{-3}$ on large
$\sim10\:$pc ``cloud'' scales to $\gtrsim 10^5\:{\rm cm}^{-3}$ in
their densest clumps and cores (e.g., P\'erault et al. 1996; Egan et
al. 1998; Rathborne et al. 2006; Simon et al. 2006; Pillai et
al. 2006; Butler \& Tan 2009, 2012; Tan et al. 2014). IRDCs exhibit
high mass surface densities ($\Sigma \sim 0.03$ to $\gtrsim 1\:{\rm
  g\:cm}^{-2}$), and their associated dust leads to high extinction,
even at mid-infrared (MIR) wavelengths.

To probe into IRDCs it is thus important to utilize far infrared (FIR)
observations. The {\it Herschel} infrared Galactic Plane (Hi-GAL) survey
(Molinari et al. 2016), is the most recent and capable FIR survey
covering large numbers of IRDCs. It provides photometric maps and
compact source catalogs at five different wavelengths: 70 $\mu$m and
160~$\mu$m using the PACS instrument; 250~$\mu$m, 350~$\mu$m, and
500~$\mu$m using the SPIRE instrument. There is also 110~$\mu$m
imaging available for certain regions from the {\it Herschel} data
archive.

Our goal in this paper is to use these {\it Herschel} data to identify
protostars and characterize their spectral energy distributions (SEDs)
in the massive, well-studied IRDC G028.37+00.07, also known as Cloud C
from the sample of Butler \& Tan (2009, 2012). Our intent is to
  develop unbiased, algorithmic methods that can eventually be
  scaled-up to much larger samples of clouds. Cloud C is located at
a kinematic distance of about 5~kpc (Simon et al. 2006) and within its
defined elliptical boundary region (of an effective radius of 7.7~pc),
it has an estimated mass of $68,300\:M_\odot$ from NIR + MIR
extinction maps (Kainulainen \& Tan 2013) and $72,000\:M_\odot$ from
an estimate of the {\it Herschel}-observed sub-mm dust emission, as
processed by Lim et al. (2016). Thus IRDC C is one of the most massive
IRDCs in the Galaxy. It appears to be a relatively coherent structure,
with a virial parameter (Bertoldi \& McKee 1992) of about unity
(Butler et al. 2014; Hernandez \& Tan 2015). This IRDC is a prime
candidate for being a massive star cluster in the early stages of its
formation.

In \S{\ref{S:methods}} we discuss our methods for identifying and
characterizing protostellar SEDs, including use of the Zhang \& Tan
(2018, hereafter ZT) radiative transfer model grid. Our paper is a
first application of these ZT models to relatively faint sources,
where uncertainties in the SEDs can be dominated by background
subtraction and include significant wavelength ranges where only upper
limits on fluxes are derived. Thus in \S{\ref{S:results}} we present
an extensive discussion of SED model fitting results and their
sensitivity to certain choices related to measuring the SEDs. We then
describe the bolometric luminosity function of the sources and the
implied protostellar mass function. We compare core envelope masses
predicted by the ZT model grid with those estimated from the commonly
used method of single temperature grey-body fitting of the SEDs. We
then consider the protostellar population as a whole, estimating the
total star formation rate, i.e., the star formation efficiency per
free-fall time, in the IRDC. Finally, we examine the clustering
properties of the sources and discuss whether there is evidence for
the most massive protostars to tend to form near the protocluster
center, i.e., primordial mass segregation, or in more clustered manner
than lower-mass sources. We discuss the implications of our results,
our general conclusions and future directions in
\S{\ref{S:conclusions}}.

\section{Methods}\label{S:methods}

\subsection{Source Identification}

\begin{figure*}
\vspace{-3.5cm}
\includegraphics[scale=0.8]{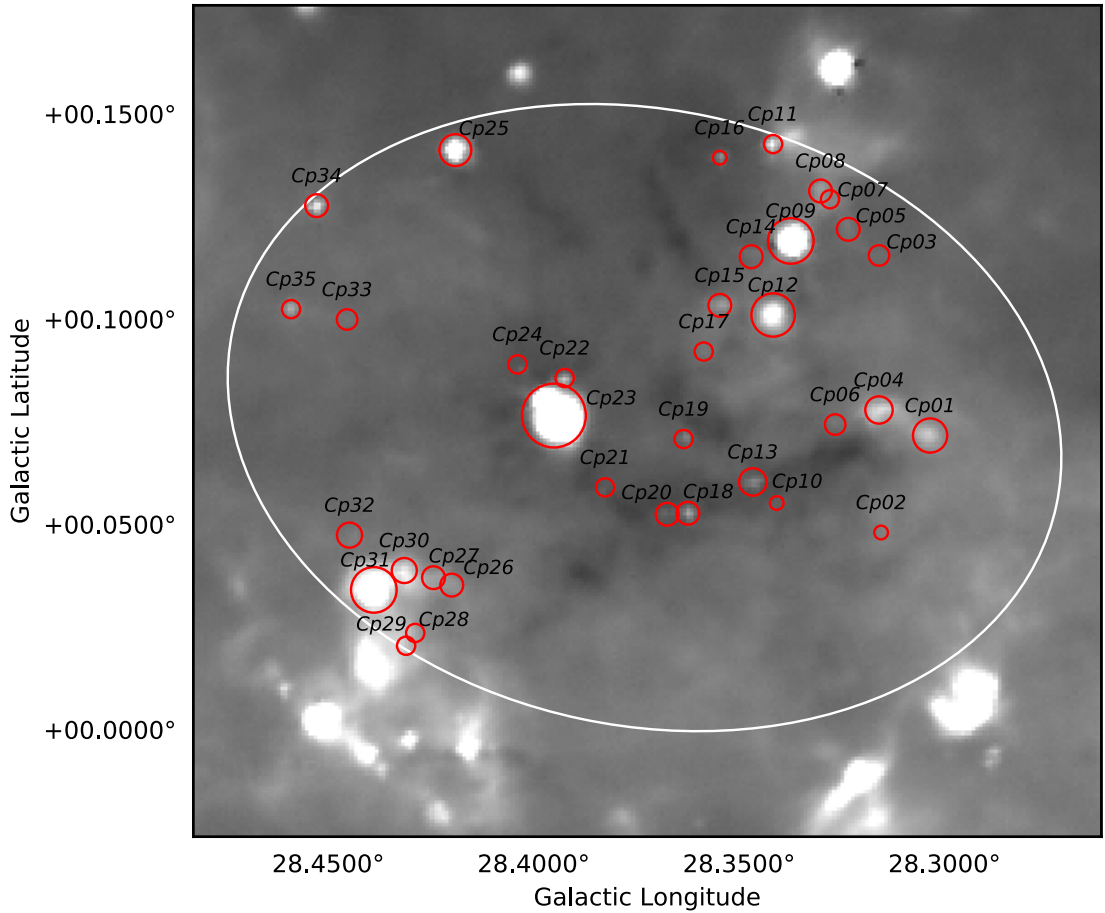}
\vspace{-5.0cm}
\caption{
Overview of IRDC G028.37+00.07 (Cloud C), showing the $70\:{\rm \mu
  m}$-identified protostellar sources, overlaid on the 70 $\mu$m
Hi-GAL image. The red circles denote the chosen aperture sizes of the
sources, described in \S\ref{S:photometry}. The white ellipse is
  the boundary of the IRDC defined by Simon et
  al. (2006).}\label{fig:map}
\end{figure*}

The catalog of Hi-GAL 70~$\rm \mu m$ point sources (Molinari et
al. 2016) forms the basis of our sample. These sources were identified
using the CUrvature Thresholding Extractor (CuTEx) algorithm, which
finds pixels with high curvature by calculating the second derivatives
in intensity profiles; areas above a certain curvature threshold
indicate the location of a source. As described by Molinari et
al. (2016), the source extraction threshold was chosen to be able to
detect relatively faint sources, while minimizing false detections.

We obtained the coordinates of all 70~$\rm \mu m$ sources in the
Hi-GAL catalog that overlap with the elliptical region defining IRDC C
(Simon et al. 2006), identifying a total of 40 sources. We then
inspected the Hi-GAL images of these sources, especially at 70~$\rm
\mu m$, to examine source crowding. We found that several sources were
in locally crowded regions, such that it is not possible to resolve
their emission at $\sim 160\:{\rm \mu m}$ near the peak of their
SEDs. The angular resolution of {\it Herschel} at these wavelengths
leads us to set a minimum aperture size of $\sim 6\arcsec$ in radius.

Thus, in these cases of source crowding we simply treat the region as
a single source. The most prominent example of source crowding is that
of Cp23, near the C9 region in Butler et al. (2014) (hereafter BTK14),
which was marked as four different sources in the Hi-GAL catalog. We
model it as a single, large source, with its coordinates chosen from
the most central of the four Hi-GAL sources. There are then only two
other cases of ``crowding'', involving close pairs of sources. Here,
we set the strongest 70~$\rm \mu m$ source to be the source location,
so all of the coordinates are still directly from the Hi-GAL
catalog. After these steps, 35 sources remain in our sample, which we
label Cp01, Cp02, Cp03, etc., i.e., protostellar candidate sources in
IRDC C, based on increasing Galactic longitude (see
Figure~\ref{fig:map}).

\subsection{Photometry and SED Construction}\label{S:photometry}

We analyzed archival 70, 110, \& 160~$\rm \mu m$ {\it Herschel}-PACS
images of proposal ID ``KPGT-okrause-1''.  These data were observed
with medium scanning speed and have 6\arcsec, 7\arcsec and 11\arcsec
angular resolution, respectively (Poglitsch et al. 2010). The data
sets were obtained as product level 2.5 and the Standard Product
Generation (SPG) v14.2.0. We applied zero-level offset correction by
following the method of Lim \& Tan (2014), which we describe below. We
adopt a model SED of the diffuse Galactic plane emission (Draine \& Li
2007) from near-infrared to sub-mm. We fit this model to the observed
median intensities from 90-110\% size annuli compared to the major and
minor axes of the IRDC that is defined by Simon et al. (2006).  We
consider data at 8~$\rm \mu m$ ({\it Spitzer}-IRAC), 24~$\rm \mu m$
({\it Spitzer}-MIPS), 70~$\rm \mu m$, 160~$\rm \mu m$ ({\it
  Herschel}-PACS), 250~$\rm \mu m$, 350~$\rm \mu m$ and 500~$\rm \mu
m$ (Herschel-SPIRE) (with these Herschel data from the Hi-GAL survey;
Molinari et al. 2010) and then predict the expected intensities in the
archival {\it Herschel}-PACS 70, 110 \& 160~$\rm \mu m$ band data.  A
single value offset for each wavelength was then applied to each
dataset (760, 2615 \& 3801 MJy/sr for the 70, 110 \& 160~$\rm \mu m$
bands, respectively). We found an astrometric difference of a few
arcseconds between the {\it Herschel} and {\it Spitzer} maps. We
corrected this by the average value of the mean positional offset of
point sources seen at 8, 24, 70, 110 and 160~$\rm \mu m$.

For the photometry of the sources at shorter wavelengths, we utilize
the 24 $\mu$m {\it Spitzer}-MIPS images from the MIPSGAL survey (Carey
et al. 2009). We also examine images from the {\it Spitzer}-IRAC
Galactic Legacy Infrared Mid-Plane Survey Extraordinaire (GLIMPSE)
(Churchwell et al 2009). Most of our sources appear ``dark'' at the
shortest IRAC wavelengths, $\sim3\:{\rm \mu m}$, and often even the
$8\:{\rm \mu m}$ image only provides an upper limit on source
flux. Given that the ZT protostellar models do not accurately predict
fluxes at these wavelengths, where PAH emission can often be
significant, we only utilize the $8\:{\rm \mu m}$ (IRAC Band 4) image
to place upper limit constraints on source SEDs.

We use fixed aperture sizes that are determined by inspecting the
morphology of the 70~$\rm \mu m$ images. The apertures were chosen 
to include as much of the emission coming from the source as possible, 
while avoiding the emission of nearby sources. Since the beam size for 
this image is 6$\arcsec$, the smallest aperture allowed for the sources 
was also set to a radius of 6$\arcsec$ in order to match the beam size. 
The majority of the sources have apertures slightly larger than the
beam size, averaging about 10$\arcsec$ in radius. We also examine the
sensitivity of our results to varying the aperture size by 30\%.
For a given aperture, then the photometric flux of each source was
measured at 8, 24, 70, 100, 160, 250, 350, and 500 $\mu$m using the
Python package PHOTUTILS. 

Since the protostellar sources are embedded in a high mass surface
density protocluster clump, it is important to carry out background
subtraction of flux from this surrounding material. We use an annular
region extending to twice the aperture radius to measure this
background emission, which follows the methods adopted previously by
De Buizer et al. (2017). The level of the background is then assessed
as the median intensity value in this annulus. We will examine
the effects on the SEDs and other results of either carrying out
(which is our fiducial case) or not carrying out this step of
background subtraction.

The uncertainties in the fluxes receive a contribution from basic
photometric/calibration uncertainties, which we set to 10\%, combined
in quadrature with those due to background subtraction, which can
often be the dominant source of uncertainty. We assess the level of
background uncertainty by examining the level of the background
fluctuations, $\sigma_{\rm bg}$, measured as the standard deviation of flux densities patches in the annular
background region that have an area equal to that of the aperture.

\newpage
\subsection{Fitting SED Models}

Once a source SED is derived, consisting of measured fluxes, including
upper limits, and their estimated uncertainties, then these data are
used to constrain the ZT protostellar SED models, under the assumption
of fixed source distances of 5~kpc. The detailed method of the fitting
procedure follows that of De Buizer et al. (2017) and Liu et
al. (2019): in particular the short wavelength IRAC 8~$\rm \mu m$ data
point is used only as an upper limit, given the uncertainties of its
possible contamination with PAH emission that is not treated in the ZT
SED models.

The physical basis of the ZT protostellar models is the Turbulent Core
Model (McKee \& Tan 2003). While there are several other grids of
protostellar radiative transfer models (e.g., Robitaille et al. 2006;
Molinari et al. 2008; Robitaille 2017), these tend to be less
physically self-consistent, especially for high pressure, high density
condition of IRDCs, and have much larger numbers of free
parameters. There are only three main parameters in the ZT models:
initial core mass, $M_c$, with the current grid exploring a range from
10 to $480\:M_\odot$; surrounding clump mass surface density,
$\Sigma_{\rm cl}$, with a range from 0.1 to 3.2$\:{\rm g\:cm}^{-2}$
(which sets the bounding pressure of the core, so cores in high
$\Sigma_{\rm cl}$ regions are smaller and denser); and the current
protostellar mass, $m_*$, which sets the evolutionary stage of the
collapse of a given core. The protostellar mass is sampled from masses
from $0.5, 1, 2... \:M_\odot$ up to masses that can be typically
$\sim50\%$ of $M_c$, with this efficiency set by protostellar outflow
feedback. In addition to these three primary parameters, the fitting
procedure also returns an estimate of the inclination angle of the
protostellar outflow axis to the line of sight and an estimate of the
foreground extinction.

\section{Results}\label{S:results}

\subsection{Examples of SED Fitting and Effects of Aperture Size}

Here we illustrate results of the SED model fitting for three example
sources: Cp23 selected as an example of a large, bright source; Cp15
as an example of a more typical source in the sample of moderate flux;
and Cp03 as an example of a relatively faint source.

\begin{deluxetable*}{ccccccccccccc}\label{tab:SEDs}
\tabletypesize{\scriptsize}
\tablecaption{Parameters of ten best ZT models for example protostars: Cp23; Cp15; Cp03}
\tablehead{
\colhead{Source} &\colhead{$\chi^{2}$} & \colhead{$M_{c}$} & \colhead{$\Sigma_{\rm cl}$} &\colhead{$R_{c}$}&\colhead{$m_{*}$} & \colhead{$\theta_{\rm view}$} &\colhead{$A_{V}$} & \colhead{$M_{\rm env}$} &\colhead{$\theta_{w,\rm esc}$} & \colhead{$\dot{m}_{*}$} & \colhead{$L_{\rm tot,iso}$} & \colhead{$L_{\rm tot,bol}$} \\
\colhead{} &\colhead{} & \colhead{($M_\odot$)} & \colhead{(g $\rm cm^{-2}$)} &\colhead{pc (\arcsec)} &\colhead{($M_{\odot}$)} & \colhead{(deg)} & \colhead{(mag)} & \colhead{($M_{\odot}$)} & \colhead{(deg)} &\colhead{($M_{\odot}$/yr)} & \colhead{($L_{\odot}$)} & \colhead{($L_{\odot}$)} \\
}
\startdata
Cp23
& 12.425 & 400 & 3.2 & 0.083 ( 3.42 ) & 8.0 & 12.84 & 36 & 382 & 7 & 1.1$\rm \times 10^{-3}$ & 4.2$\rm \times 10^{4}$ & 2.0$\rm \times 10^{4}$ \\
$R_{ap}$ = 28 \arcsec
& 13.171 & 320 & 3.2 & 0.074 ( 3.05 ) & 8.0 & 12.84 & 55.6 & 308 & 8 & 1.0$\rm \times 10^{-3}$ & 4.4$\rm \times 10^{4}$ & 1.7$\rm \times 10^{4}$ \\
& 14.311 & 480 & 3.2 & 0.091 ( 3.74 ) & 8.0 & 12.84 & 17.2 & 462 & 6 & 1.1$\rm \times 10^{-3}$ & 4.0$\rm \times 10^{4}$ & 2.2$\rm \times 10^{4}$ \\
& 17.961 & 240 & 3.2 & 0.064 ( 2.65 ) & 8.0 & 12.84 & 87.9 & 227 & 10 & 9.5$\rm \times 10^{-4}$ & 6.0$\rm \times 10^{4}$ & 1.7$\rm \times 10^{4}$ \\
& 22.821 & 200 & 3.2 & 0.059 ( 2.42 ) & 8.0 & 12.84 & 100.0 & 184 & 11 & 9.0$\rm \times 10^{-4}$ & 7.8$\rm \times 10^{4}$ & 2.0$\rm \times 10^{4}$ \\
& 28.397 & 200 & 3.2 & 0.059 ( 2.42 ) & 4.0 & 12.84 & 7.1 & 191 & 7 & 6.5$\rm \times 10^{-4}$ & 2.5$\rm \times 10^{4}$ & 1.2$\rm \times 10^{4}$ \\
& 29.639 & 160 & 3.2 & 0.052 ( 2.16 ) & 8.0 & 22.33 & 0.0 & 146 & 13 & 8.5$\rm \times 10^{-4}$ & 1.9$\rm \times 10^{4}$ & 2.0$\rm \times 10^{4}$ \\
& 29.948 & 400 & 1.0 & 0.147 ( 6.07 ) & 8.0 & 12.84 & 0.0 & 383 & 8 & 4.6$\rm \times 10^{-4}$ & 2.3$\rm \times 10^{4}$ & 1.2$\rm \times 10^{4}$ \\
& 31.283 & 480 & 0.3 & 0.287 ( 11.83 ) & 12.0 & 22.33 & 72.7 & 459 & 10 & 2.5$\rm \times 10^{-4}$ & 3.8$\rm \times 10^{4}$ & 4.0$\rm \times 10^{4}$ \\
& 32.020 & 480 & 1.0 & 0.161 ( 6.65 ) & 12.0 & 12.84 & 100.0 & 461 & 9 & 5.9$\rm \times 10^{-4}$ & 8.5$\rm \times 10^{4}$ & 3.8$\rm \times 10^{4}$ \\
Averages
& 15.720 & 312 & 3.2 & 0.073 ( 3.02 ) & 8.0 & 12.84 & 59.4 & 296 & 9 & 1.0$\rm \times 10^{-3}$ & 5.1$\rm \times 10^{4}$ & 2.0$\rm \times 10^{4}$ \\
\hline\noalign{\smallskip}
Cp15
& 0.662 & 80 & 0.1 & 0.208 ( 8.59 ) & 1.0 & 88.57 & 100 & 77 & 8 & 1.9$\rm \times 10^{-5}$ & 1.7$\rm \times 10^{2}$ & 1.9$\rm \times 10^{2}$ \\
$R_{ap}$ = 10 \arcsec
& 0.662 & 60 & 0.1 & 0.180 ( 7.44 ) & 1.0 & 88.57 & 100.0 & 57 & 10 & 1.8$\rm \times 10^{-5}$ & 1.7$\rm \times 10^{2}$ & 2.0$\rm \times 10^{2}$ \\
& 0.849 & 50 & 0.1 & 0.165 ( 6.79 ) & 1.0 & 61.64 & 100.0 & 48 & 11 & 1.7$\rm \times 10^{-5}$ & 1.5$\rm \times 10^{2}$ & 1.7$\rm \times 10^{2}$ \\
& 0.906 & 200 & 0.1 & 0.329 ( 13.58 ) & 0.5 & 12.84 & 48.5 & 200 & 3 & 1.7$\rm \times 10^{-5}$ & 1.5$\rm \times 10^{2}$ & 1.3$\rm \times 10^{2}$ \\
& 0.910 & 100 & 0.1 & 0.233 ( 9.60 ) & 1.0 & 88.57 & 100.0 & 98 & 7 & 2.0$\rm \times 10^{-5}$ & 1.8$\rm \times 10^{2}$ & 2.0$\rm \times 10^{2}$ \\
& 1.024 & 10 & 0.3 & 0.041 ( 1.71 ) & 0.5 & 28.96 & 78.8 & 9 & 18 & 1.9$\rm \times 10^{-5}$ & 1.4$\rm \times 10^{2}$ & 1.9$\rm \times 10^{2}$ \\
& 1.065 & 40 & 0.1 & 0.147 ( 6.07 ) & 2.0 & 88.57 & 100.0 & 36 & 19 & 2.2$\rm \times 10^{-5}$ & 1.8$\rm \times 10^{2}$ & 2.7$\rm \times 10^{2}$ \\
& 1.211 & 40 & 0.1 & 0.147 ( 6.07 ) & 1.0 & 54.90 & 100.0 & 38 & 12 & 1.6$\rm \times 10^{-5}$ & 1.4$\rm \times 10^{2}$ & 1.7$\rm \times 10^{2}$ \\
& 1.407 & 160 & 0.1 & 0.294 ( 12.14 ) & 0.5 & 22.33 & 0.0 & 158 & 3 & 1.6$\rm \times 10^{-5}$ & 1.0$\rm \times 10^{2}$ & 9.8$\rm \times 10^{1}$ \\
& 1.435 & 200 & 0.1 & 0.329 ( 13.58 ) & 1.0 & 85.70 & 84.8 & 197 & 4 & 2.5$\rm \times 10^{-5}$ & 1.7$\rm \times 10^{2}$ & 1.8$\rm \times 10^{2}$ \\
Averages
& 0.980 & 69 & 0.1 & 0.183 ( 7.53 ) & 0.9 & 62.06 & 81.2 & 66 & 10 & 1.9$\rm \times 10^{-5}$ & 1.5$\rm \times 10^{2}$ & 1.8$\rm \times 10^{2}$ \\
\hline\noalign{\smallskip}
Cp03
& 0.248 & 10 & 0.3 & 0.041 ( 1.71 ) & 4.0 & 77.00 & 100 & 1 & 68 & 2.4$\rm \times 10^{-5}$ & 4.9$\rm \times 10^{1}$ & 6.7$\rm \times 10^{2}$ \\
$R_{ap}$ = 9 \arcsec
& 0.279 & 10 & 0.1 & 0.074 ( 3.04 ) & 2.0 & 79.92 & 5.1 & 4 & 50 & 1.1$\rm \times 10^{-5}$ & 2.1$\rm \times 10^{1}$ & 1.3$\rm \times 10^{2}$ \\
& 0.331 & 40 & 0.1 & 0.147 ( 6.07 ) & 12.0 & 88.57 & 100.0 & 2 & 82 & 9.5$\rm \times 10^{-6}$ & 5.7$\rm \times 10^{1}$ & 1.1$\rm \times 10^{4}$ \\
& 0.388 & 30 & 0.3 & 0.072 ( 2.96 ) & 12.0 & 88.57 & 100.0 & 1 & 81 & 2.2$\rm \times 10^{-5}$ & 7.0$\rm \times 10^{1}$ & 1.2$\rm \times 10^{4}$ \\
& 0.539 & 10 & 0.1 & 0.074 ( 3.04 ) & 1.0 & 88.57 & 100.0 & 7 & 31 & 1.0$\rm \times 10^{-5}$ & 4.4$\rm \times 10^{1}$ & 1.1$\rm \times 10^{2}$ \\
& 0.838 & 10 & 0.1 & 0.074 ( 3.04 ) & 0.5 & 88.57 & 100.0 & 9 & 20 & 7.8$\rm \times 10^{-6}$ & 4.6$\rm \times 10^{1}$ & 7.5$\rm \times 10^{1}$ \\
& 3.444 & 10 & 0.3 & 0.041 ( 1.71 ) & 2.0 & 88.57 & 100.0 & 5 & 43 & 3.0$\rm \times 10^{-5}$ & 5.8$\rm \times 10^{1}$ & 2.8$\rm \times 10^{2}$ \\
& 4.003 & 20 & 0.1 & 0.104 ( 4.29 ) & 0.5 & 88.57 & 100.0 & 19 & 13 & 9.6$\rm \times 10^{-6}$ & 7.0$\rm \times 10^{1}$ & 9.0$\rm \times 10^{1}$ \\
& 5.088 & 30 & 0.1 & 0.127 ( 5.26 ) & 0.5 & 88.57 & 100.0 & 29 & 10 & 1.1$\rm \times 10^{-5}$ & 7.6$\rm \times 10^{1}$ & 9.0$\rm \times 10^{1}$ \\
& 5.484 & 40 & 0.1 & 0.147 ( 6.07 ) & 0.5 & 88.57 & 100.0 & 39 & 8 & 1.1$\rm \times 10^{-5}$ & 7.8$\rm \times 10^{1}$ & 8.8$\rm \times 10^{1}$ \\
Averages
& 0.399 & 15 & 0.1 & 0.075 ( 3.08 ) & 2.9 & 85.20 & 84.2 & 3 & 55 & 1.3$\rm \times 10^{-5}$ & 4.5$\rm \times 10^{1}$ & 3.4$\rm \times 10^{2}$ \\
\enddata
\tablenotetext{}{The ten best models of the three example sources with the eleventh line for each source being the calculated average of ``good'' models (see text), using the fiducial aperture with background subtraction.}
\end{deluxetable*}

\begin{figure*}
\gridline{\fig{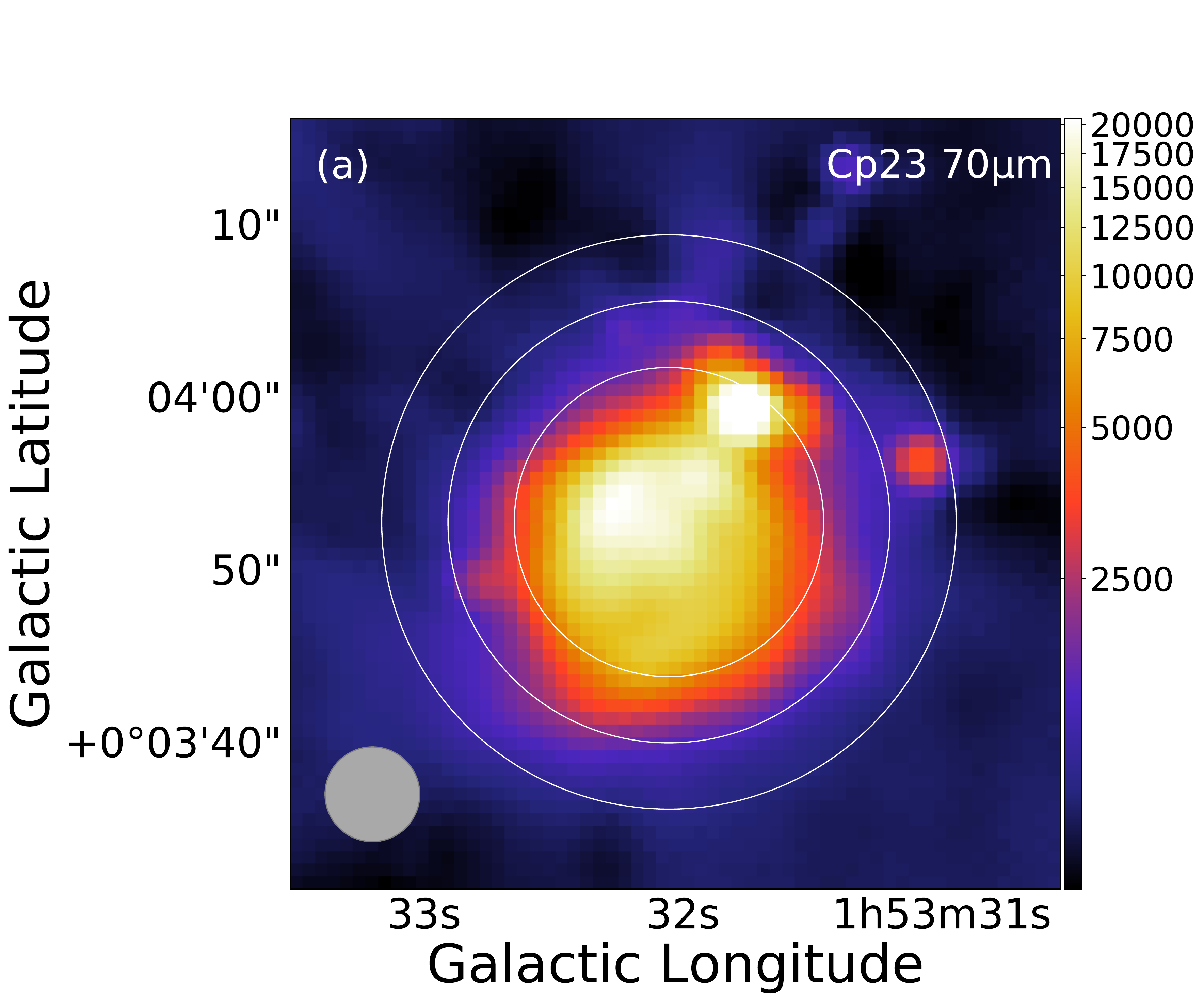}{0.231\textwidth}{}
	\hspace*{-0.95cm}
          \fig{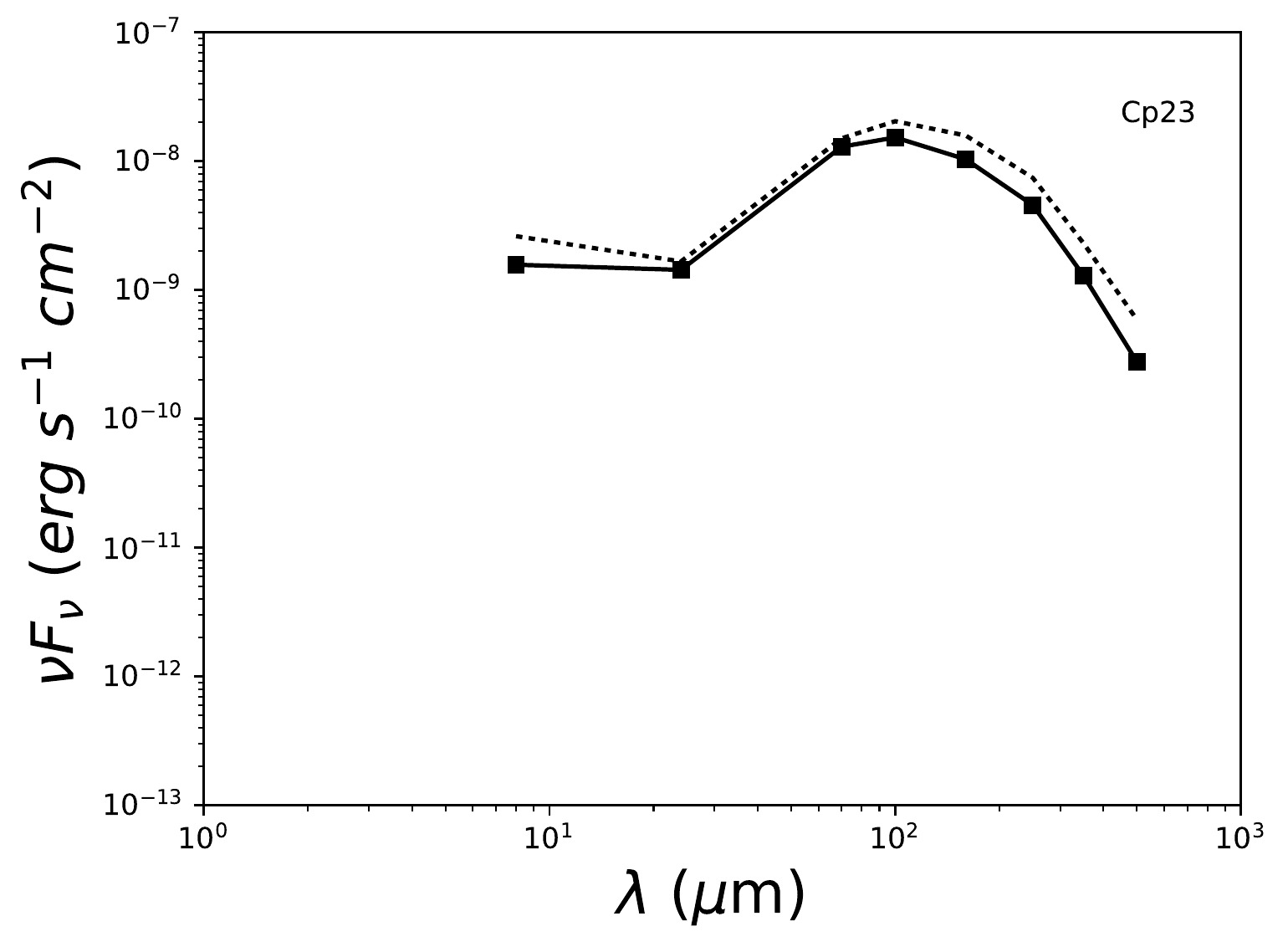}{0.245\textwidth}{}
          \hspace*{-1cm}
          \fig{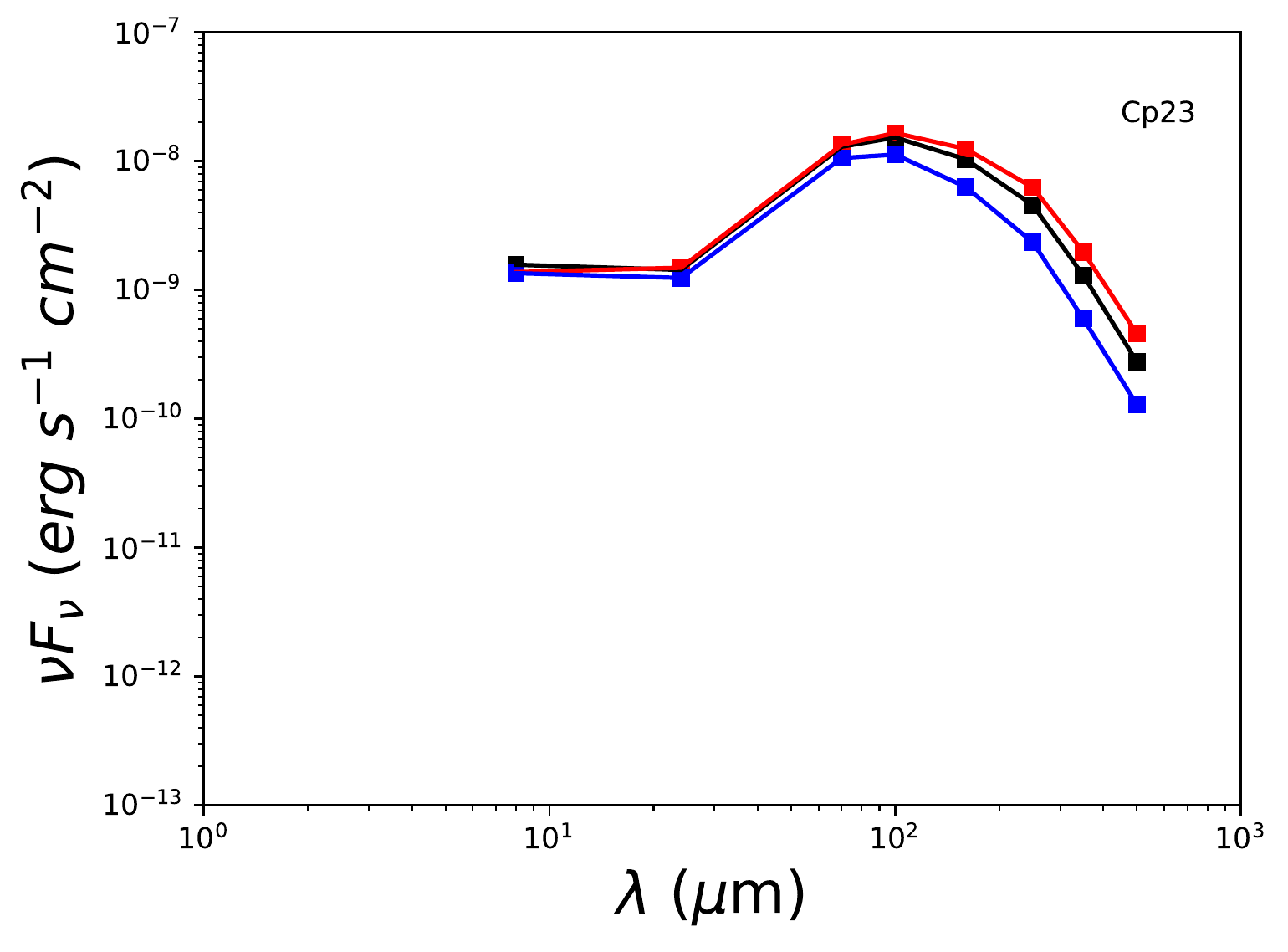}{0.245\textwidth}{}
          \hspace*{-1.1cm}
          \fig{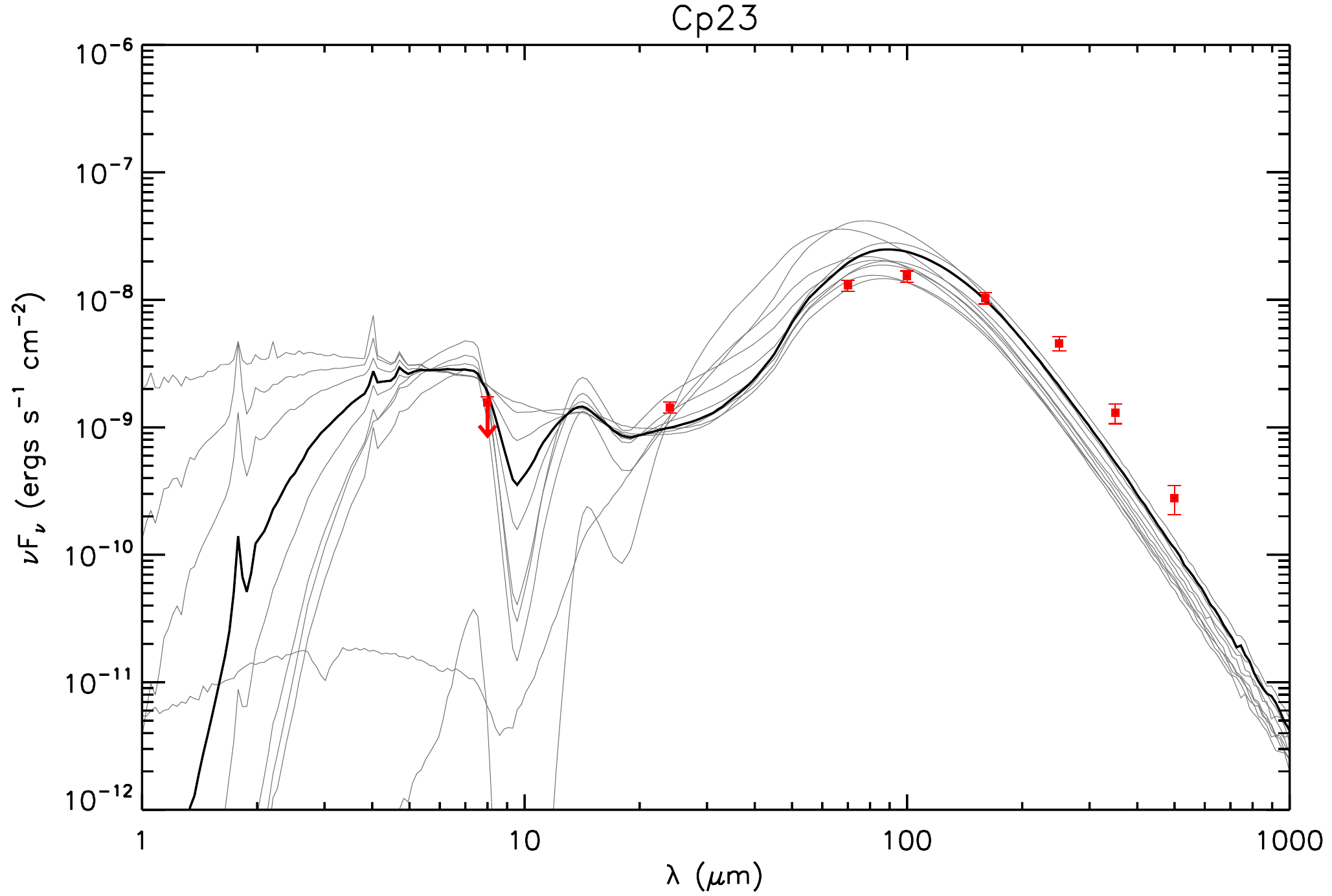}{0.264\textwidth}{}
          }
\vspace*{-0.95cm}
\gridline{\fig{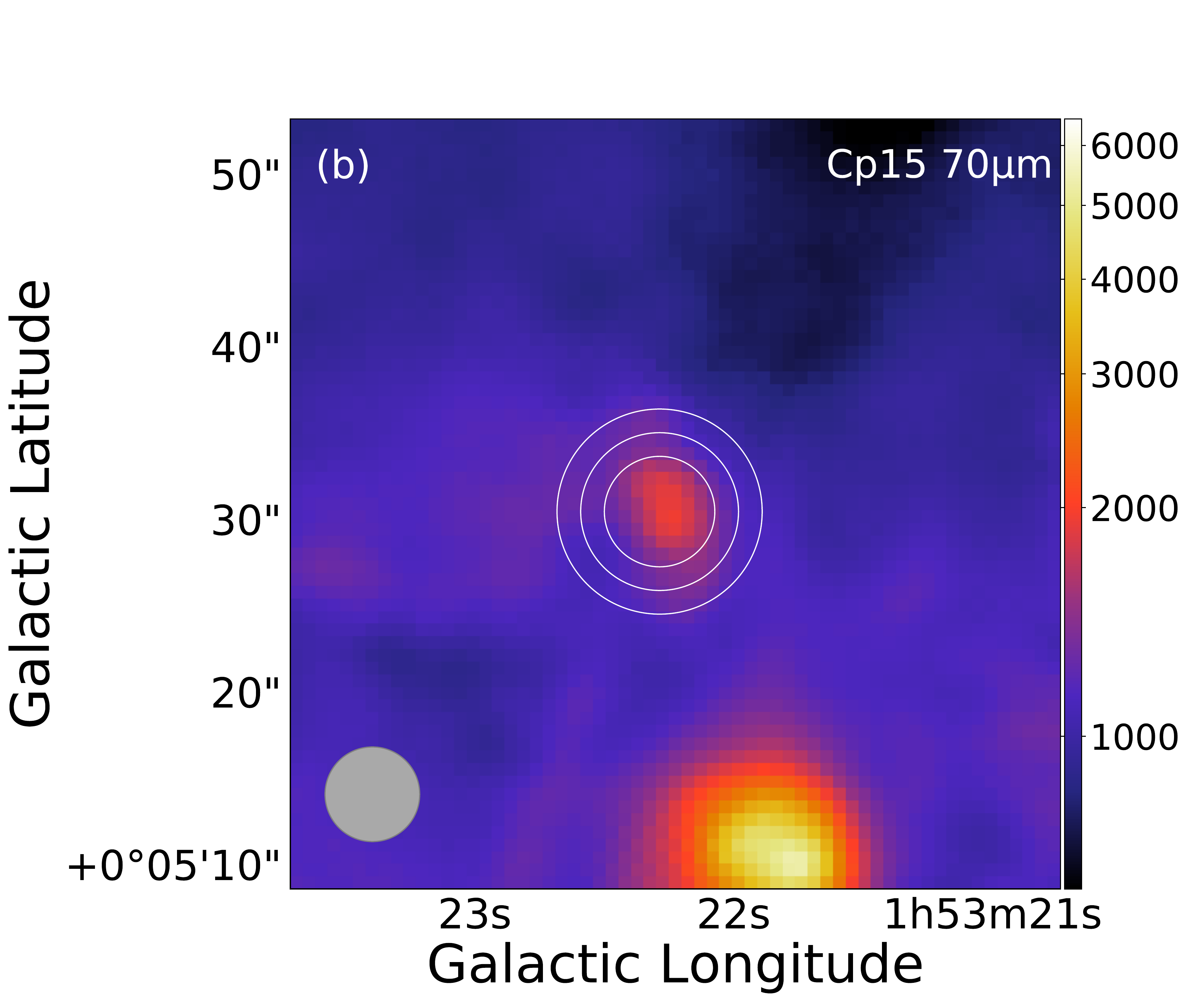}{0.231\textwidth}{}
	\hspace*{-0.95cm}
	\fig{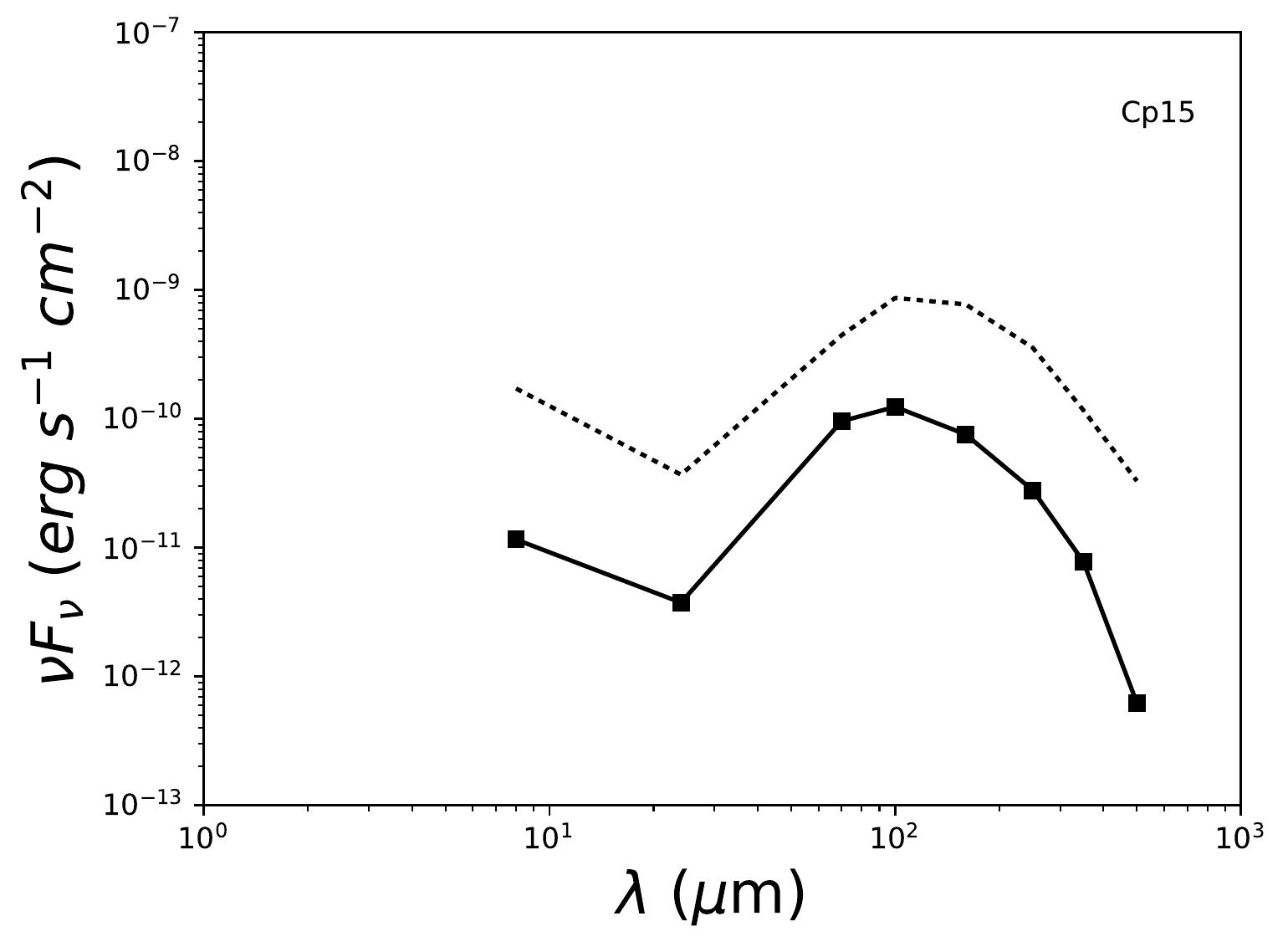}{0.245\textwidth}{}
	\hspace*{-1cm}
	\fig{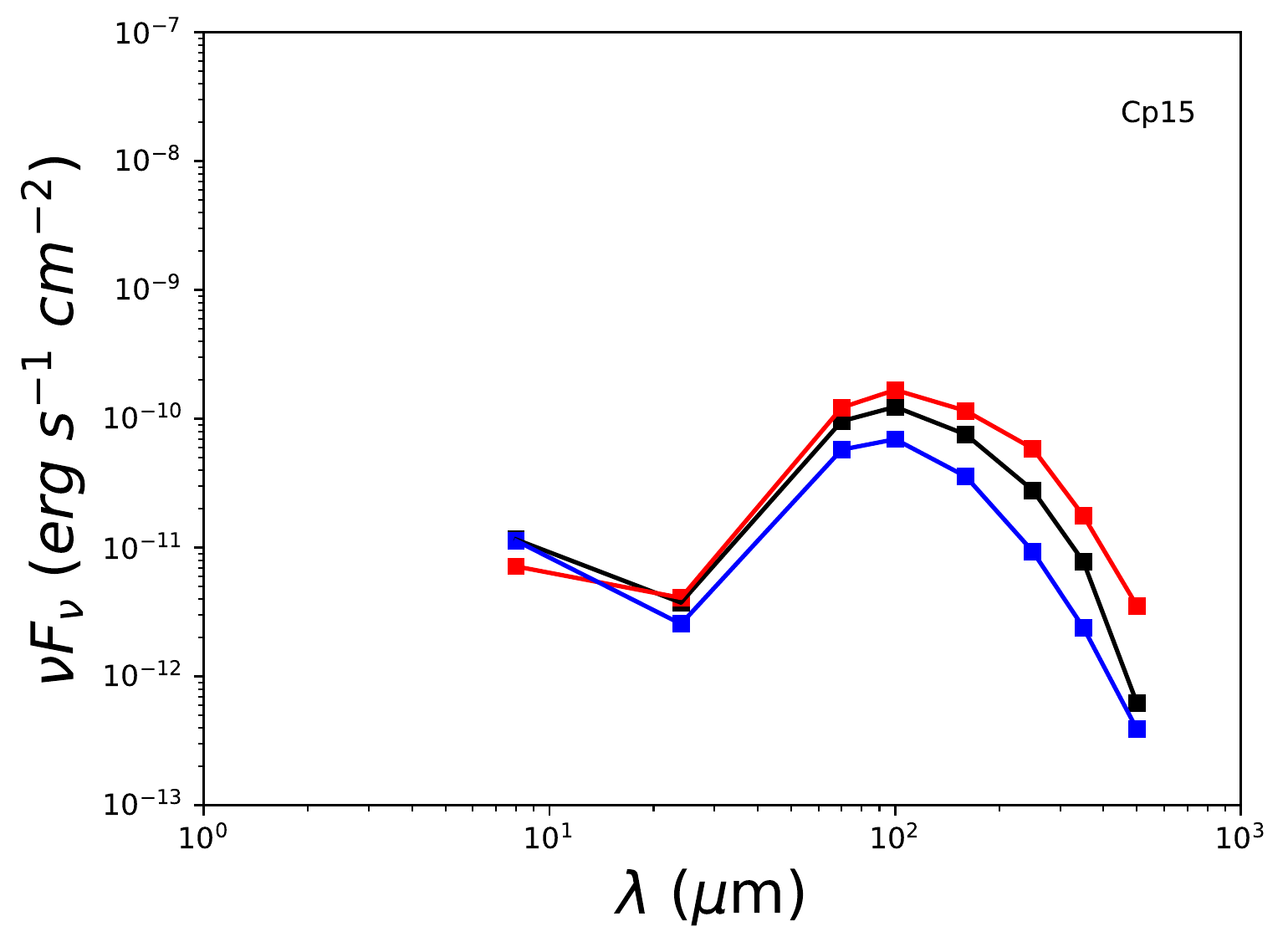}{0.245\textwidth}{}
	\hspace*{-1.1cm}
	\fig{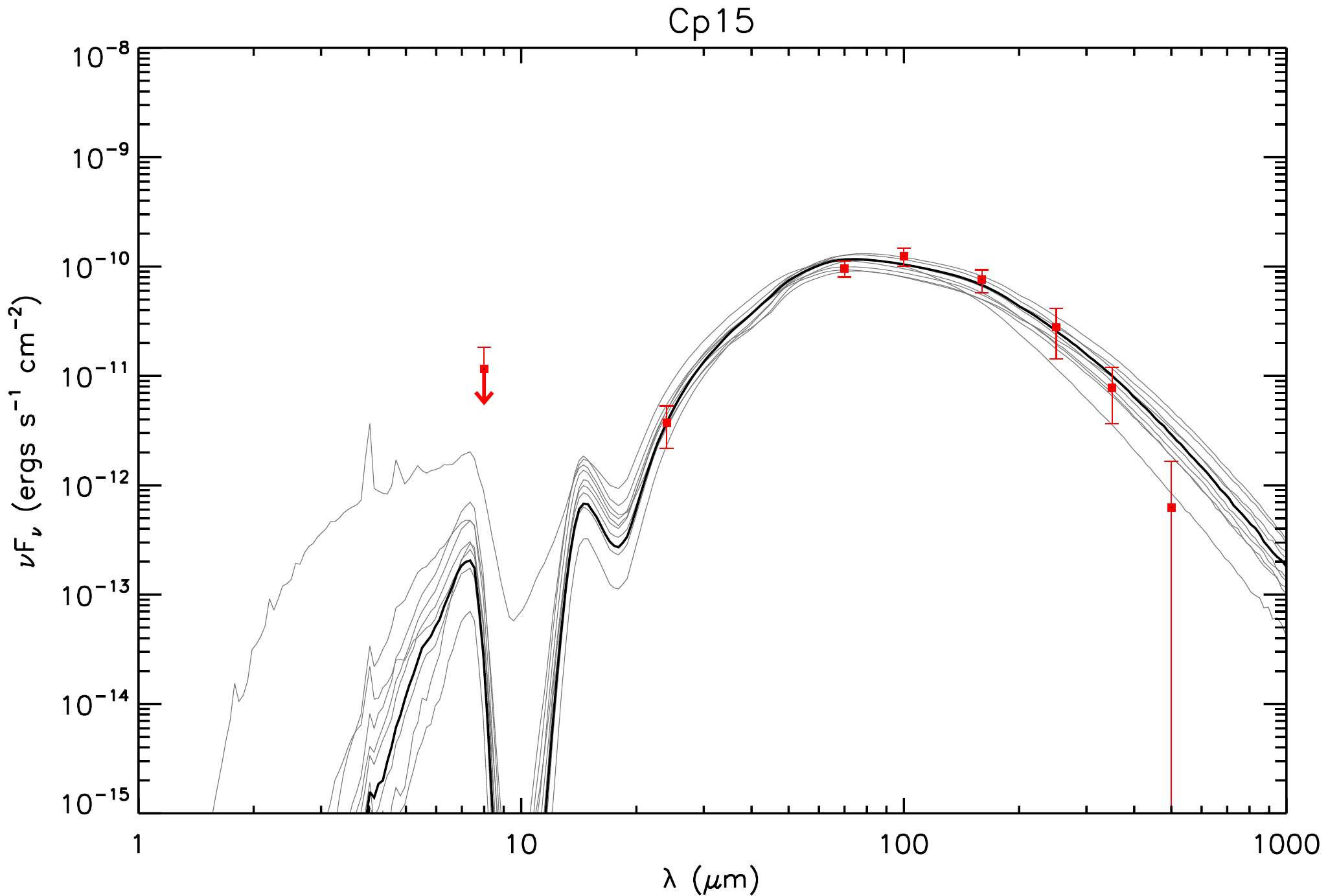}{0.264\textwidth}{}
	}
\vspace*{-0.95cm}
\gridline{\fig{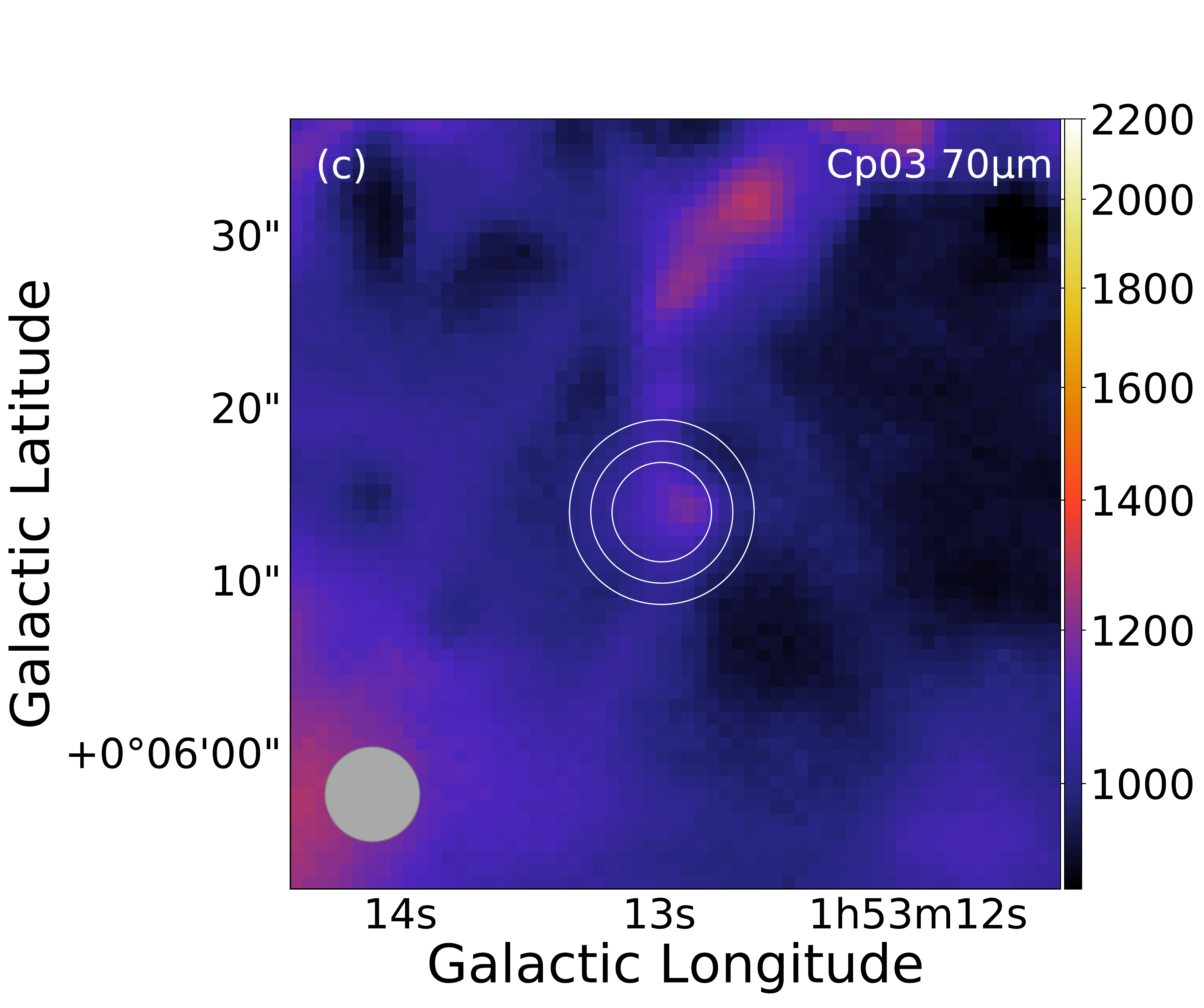}{0.231\textwidth}{}
	\hspace*{-0.95cm}
	\fig{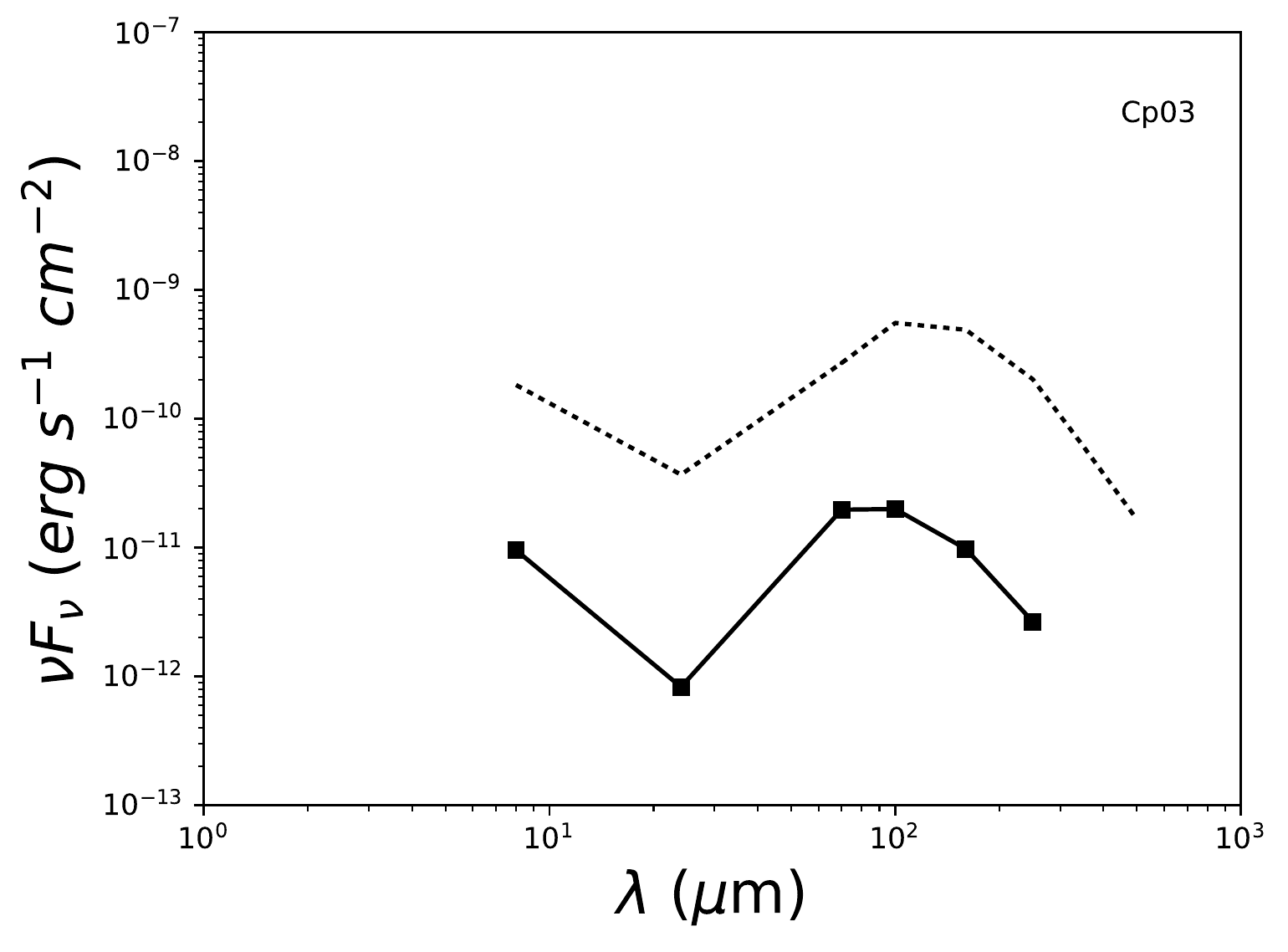}{0.245\textwidth}{}
	\hspace*{-1cm}
	\fig{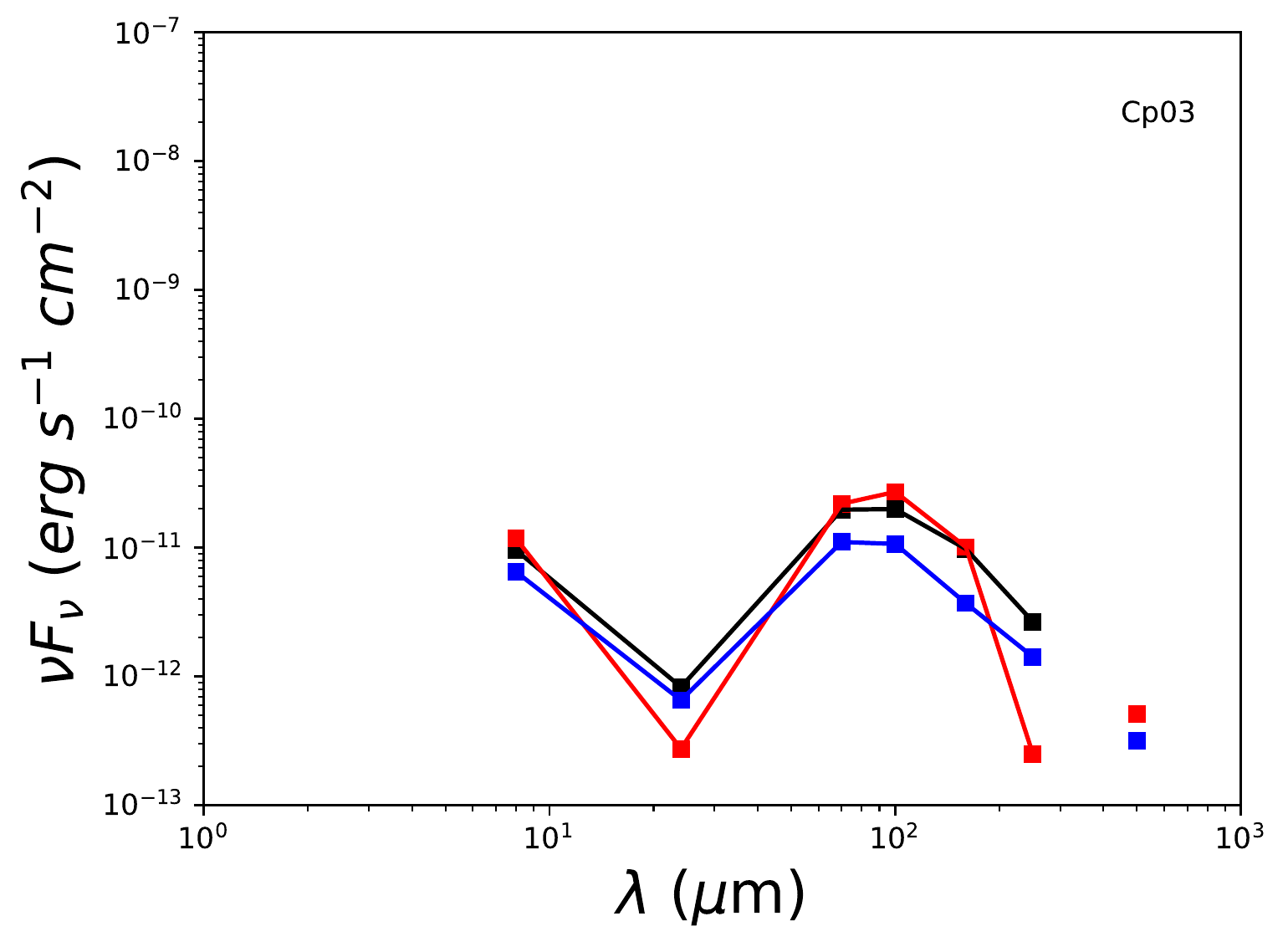}{0.245\textwidth}{}
	\hspace*{-1.1cm}
	\fig{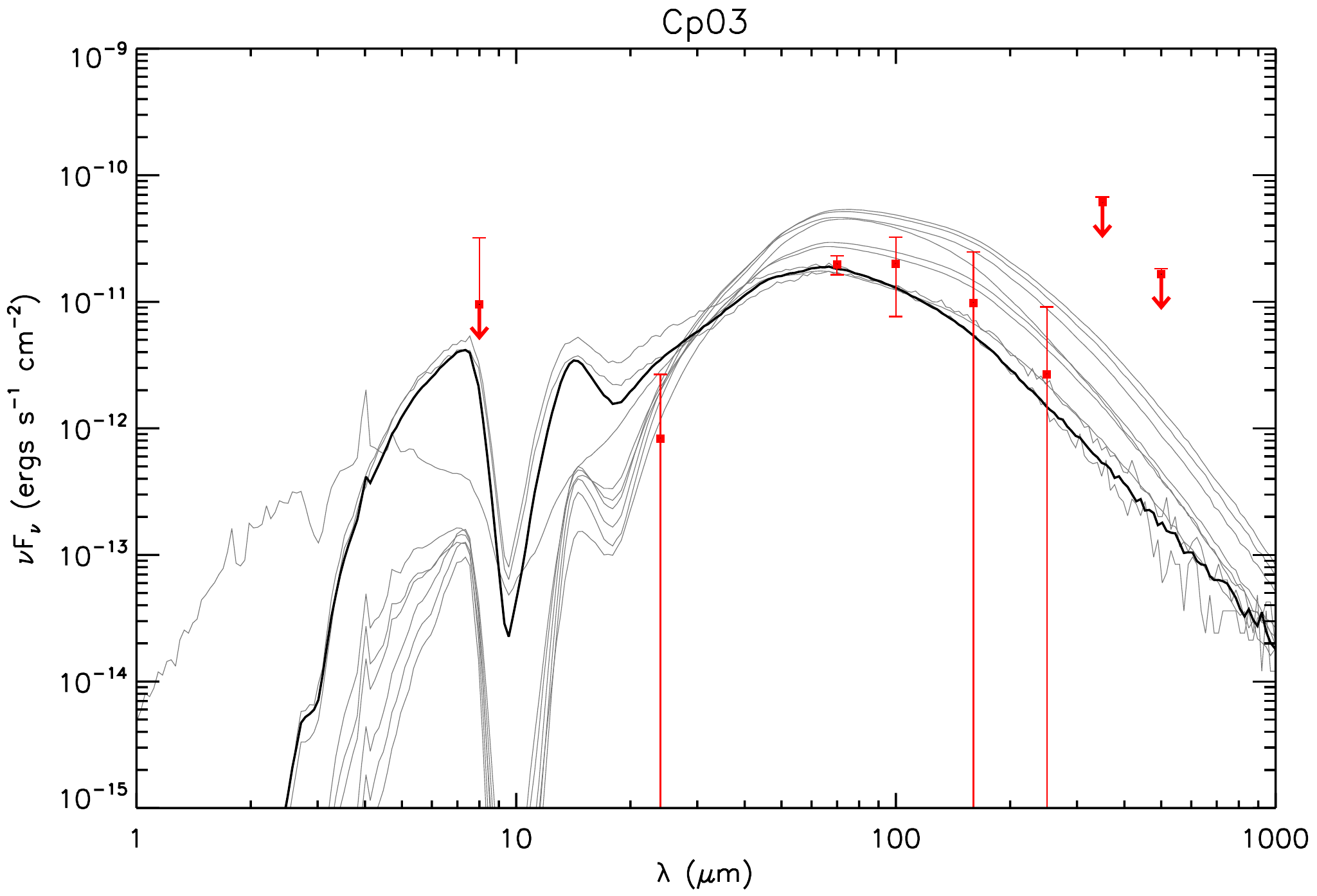}{0.264\textwidth}{}
	}
\vspace{-0.5cm}
\caption{Example protostellar sources and SED fitting: bright source Cp23 (top
row); moderate source Cp15 (middle row); and faint source Cp03 (bottom
row). The first column shows the 70~$\rm \mu m$ {\it Herschel} PACS
images of the sources, including fiducial aperture sizes (middle white
circles) and small/large apertures (inner/outer white circles). The
angular resolution of the images is shown with a filled grey circle in
lower left corner. The second column shows source SEDs (data points;
note these are simply connected by straight lines) based on the
fiducial aperture both before background subtraction (dashed lines)
and after background subtraction (solid lines). The third column shows
the effect on the background subtracted SEDs from varying the aperture
sizes, i.e., by 30\% smaller (blue) or larger (red) compared to the
fiducial (black). The fourth column shows the results of fitting ZT
protostellar SED models to the fiducial background subtracted SEDs,
with the ten best models shown (heavy line is the best model; see
text).}\label{fig:examples}
\end{figure*}

The left column of Figure~\ref{fig:examples} shows the 70~$\rm \mu m$
images of Cp23, Cp15 and Cp03 (top, middle, bottom), along with the
fiducial choice of aperture for each source (middle circles). The
second column then shows the derived source SEDs, both before
background subtraction (dashed lines) and after (solid lines). One can
see that background subtraction has a much greater effect for
the fainter sources. The third column shows the effect of different
aperture sizes, varying the radius by 30\% to smaller and larger
sizes, on the background subtracted SEDs. Finally, the fourth column
shows the data for the fiducial SEDs and the ZT model fits to these
data. Note, for Cp03 the two longest wavelength measurements for the
SED become negative after background subtraction in the fiducial case
and at these wavelengths we assume upper limits during the model
fitting process, with the values set by the level of background
fluctuations.

The parameters of the ten best SED models for each source are shown in
Table 1 in order of decreasing goodness of fit (i.e.,
increasing value of reduced $\chi^2$ [this is normalized by the number
  of data points, $N$]; 2nd column). The presentation here follows a
similar format as that of De Buizer et al. (2017) for the fitting
results of eight massive protostars in the SOMA Survey. The other
parameters presented are: initial core mass, $M_c$; mass surface
density of the clump environment, $\Sigma_{\rm cl}$; initial core
radius, $R_c$, which is determined by $M_c$ and $\Sigma_{\rm
    cl}$ and is listed in both parsecs and angular size that can be
compared to the aperture size; the current protostellar mass, $m_*$;
the viewing angle to the outflow axis, $\theta_{\rm view}$; the
foreground extinction, $A_V$; the current remaining gas mass in the
infall envelope, $M_{\rm env}$, i.e., given what has been
  accreted and expelled by feedback; the opening angle of the outflow
cavity, $\theta_{\rm w,esc}$; the accretion rate of the star,
$\dot{m}_*$; the total luminosity of the source assuming isotropic
emission given the received bolometric flux from the model,
$L_{\rm tot,iso}$; and the actual total bolometric luminosity of the
protostar model, $L_{\rm tot,bol}$.

The last row for each source in Table 1 displays the
average of each listed parameter for ``good'' model fits, using the
following method.  We have two different methods based on the
distribution of $\chi^2$ values. The first is for sources such as Cp23
that have all values of $\chi^2$ greater than 1. Here, we take the
geometric mean of the parameters of all the models with $\chi^2$ less than or equal
to twice the first, i.e., smallest, value of $\chi^2$ value. This acts
to exclude models with relatively high $\chi^2$. For example, the
average for Cp23 would include all of the models with
$\chi^2\leq2\times12.425$, which are the top five models. The second
method is for sources like Cp15 and Cp03 which have a best $\chi^2$
value smaller than 1. Here we set a limit of $\chi^2<2$,
and take the geometric mean of all the values of models from the best
set of models, up to ten, that meet this limit.

For Cp23, the best-fit model has $\chi^2$ = 12.425, which is a
relatively large value, i.e., the models do not fit particularly
well. We discuss the reasons for this below. Still, considering the
properties of the best model, we see it has an initial core mass of
$M_c=400\:M_{\odot}$, current protostellar mass of $m_*=8\:M_{\odot}$,
forming in a clump mass surface density of $\Sigma_{\rm cl}=3.2\: {\rm
  g cm}^{-2}$, and a total luminosity of $2\times10^4\:L_\odot$.
The range of values of these parameters of the best models do not vary
greatly, with the averages of ``good'' models being $M_c=312
M_{\odot}$, $\Sigma_{\rm cl} = 3.2\:{\rm g\:cm}^{-2}$ and $m_*=8\:
M_{\odot}$.

\begin{figure*}[t]
  \includegraphics[scale=0.6]{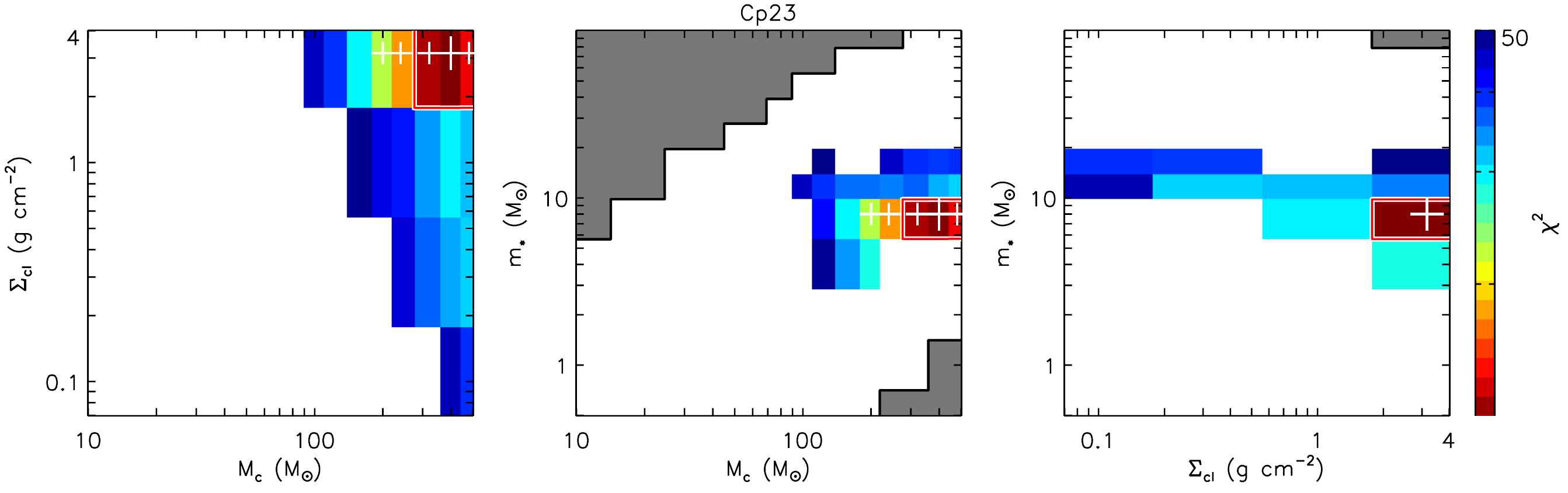}
  \includegraphics[scale=0.6]{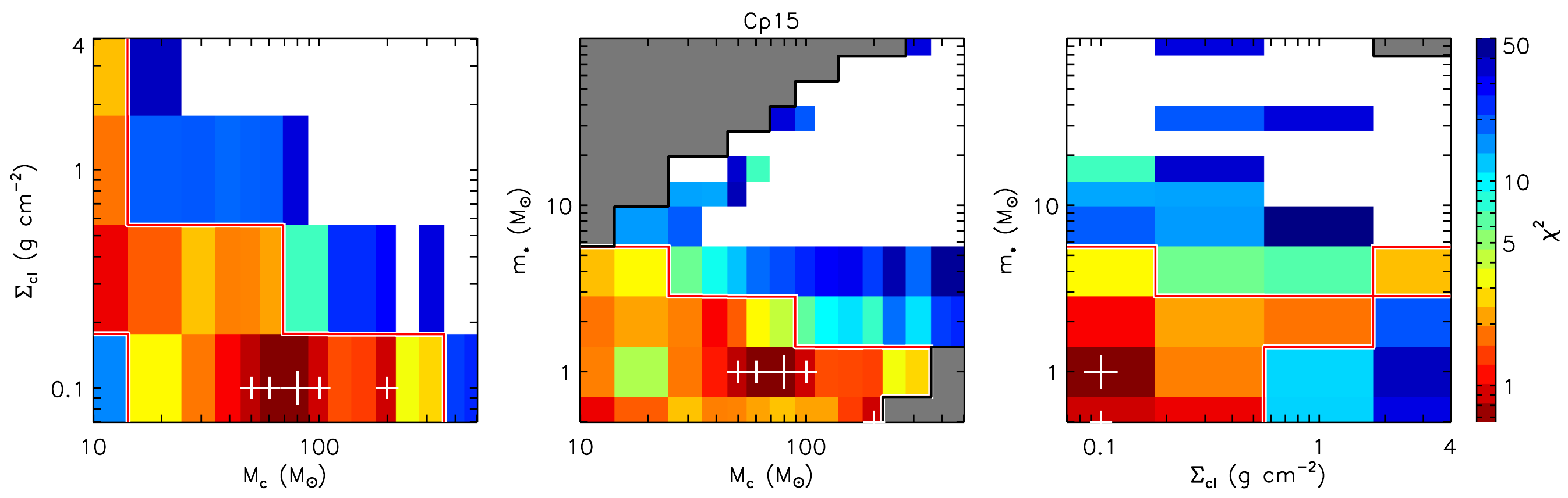}
  \includegraphics[scale=0.6]{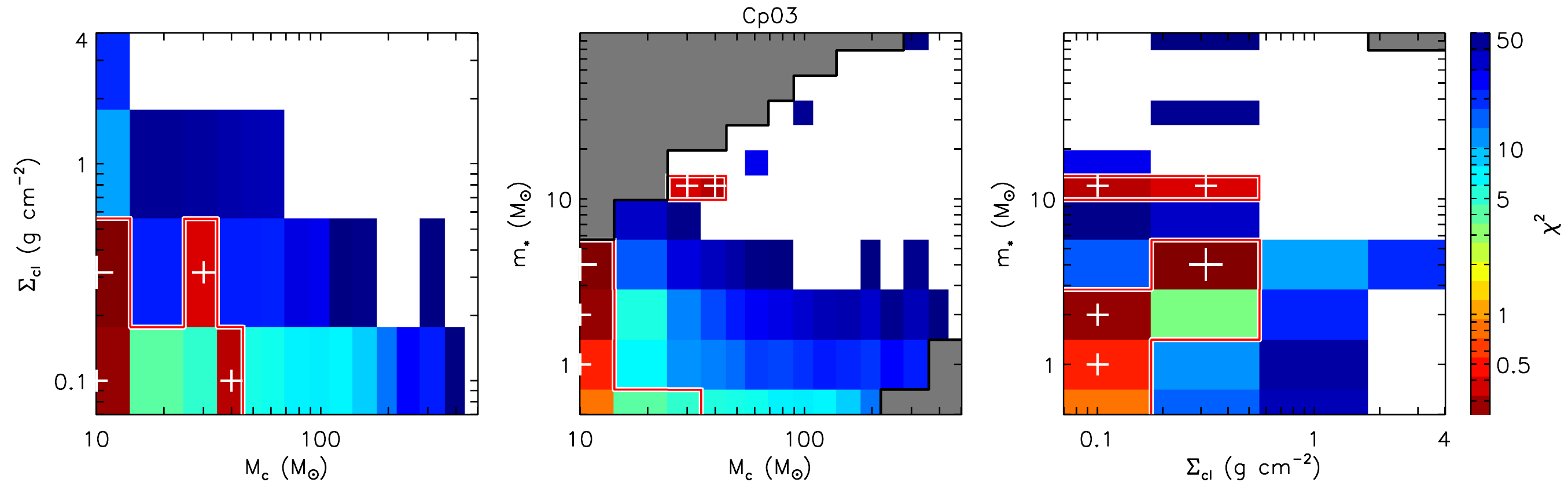}
  \caption{
Model parameter space constraints for Cp23 (a: top row), Cp15 (b:
middle row) and Cp03 (c: bottom row), following a standard output
format from the ZT model grid fitting routines. The three primary
parameters of initial core mass, $M_c$, clump mass surface density,
$\Sigma_{\rm cl}$, and current protostellar mass, $m_*$, are shown
with the color indicating reduced $\chi^2$ (white areas are models
with $\chi^2 > 50$). The best five models are shown with crosses, with
the very best model having a large cross. Gray areas are outside of the
range that is covered by the ZT model grid: especially the efficiency
of star formation from a core limits the space in the upper left of
the $m_*$ versus $M_c$ diagram. The red contours are at the level of
$\chi^2=\chi_{\rm min}^2+5$.}
\label{fig:Cp23}
\end{figure*}

A more complete view of the model parameter space for Cp23 is shown in
Figure~\ref{fig:Cp23}a, which is a standard output of the ZT model
fitting routines. The figure shows a series of 2D parameter space
plots that illustrate all the models with $\chi^2 < 50$ and with the
best five models shown with crosses (the best model has a large
cross). These plots show the correlations and degeneracies in the
resulting model parameters that are constrained by the SED data.

For Cp15, which has its model parameter space displayed in
Figure~\ref{fig:Cp23}b, we see that the preferred models shift to lower
core ($\lesssim 100\:M_\odot$) and protostellar ($\sim1\:M_\odot$)
masses. Lower clump mass surface densities also tend to be
selected. There is somewhat great dispersion in certain parameters,
such as $M_c$ and $\Sigma_{\rm cl}$, i.e., they are not as tightly
constrained as in the case of Cp23.

These trends continue for Cp03, which has its model parameter space
displayed in Figure~\ref{fig:Cp23}c. However, now we also see the
models with lowest core mass, i.e., $M_c=10\:M_\odot$, are quite
strongly preferred, though not exclusively. Such values are at the
lower boundary of the current model grid parameter space, so caution
is needed in the interpretation of the results. In particular, it is
possible that lower core masses could be reasonable fits to the data.

Next we investigate the effects of not carrying out background
subtraction and of varying aperture size when background subtraction
is carried out, on the model fitting results. Table 2 shows the values
of $\chi^2$ and various model parameters of the best fitting models and
the average of ``good'' models (see above) for these cases. Focusing
on average values, we see the general reduction of core mass, envelope
mass and luminosity following background subtraction, with relatively
larger effects seen for the lower luminosity sources Cp15 and Cp03
compared to Cp23. We also see the expected dependence of derived model
properties on aperture size, i.e., smaller masses and luminosities
when smaller apertures are used. The ranges in these values gives some
guidance on the degree of systematic uncertainties that result from
the process of background subtraction and choice of aperture
size. Note, the size of these uncertainties depends on the source
luminosity.

\newpage
\subsection{Protostellar Luminosity and Mass Functions}

\begin{figure*}
\gridline{\fig{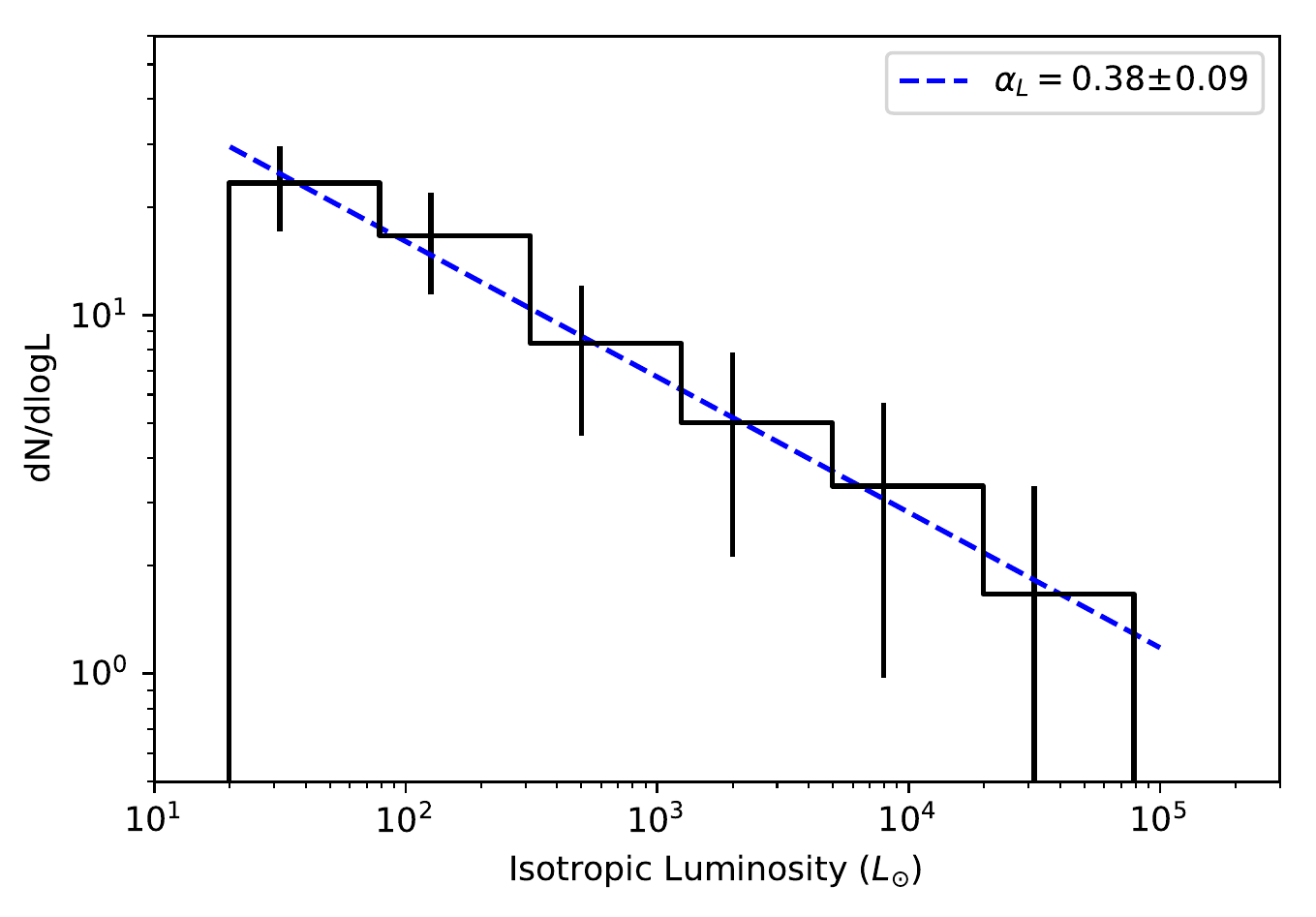}{0.45\textwidth}{}
	\hspace*{-1.6cm}
          \fig{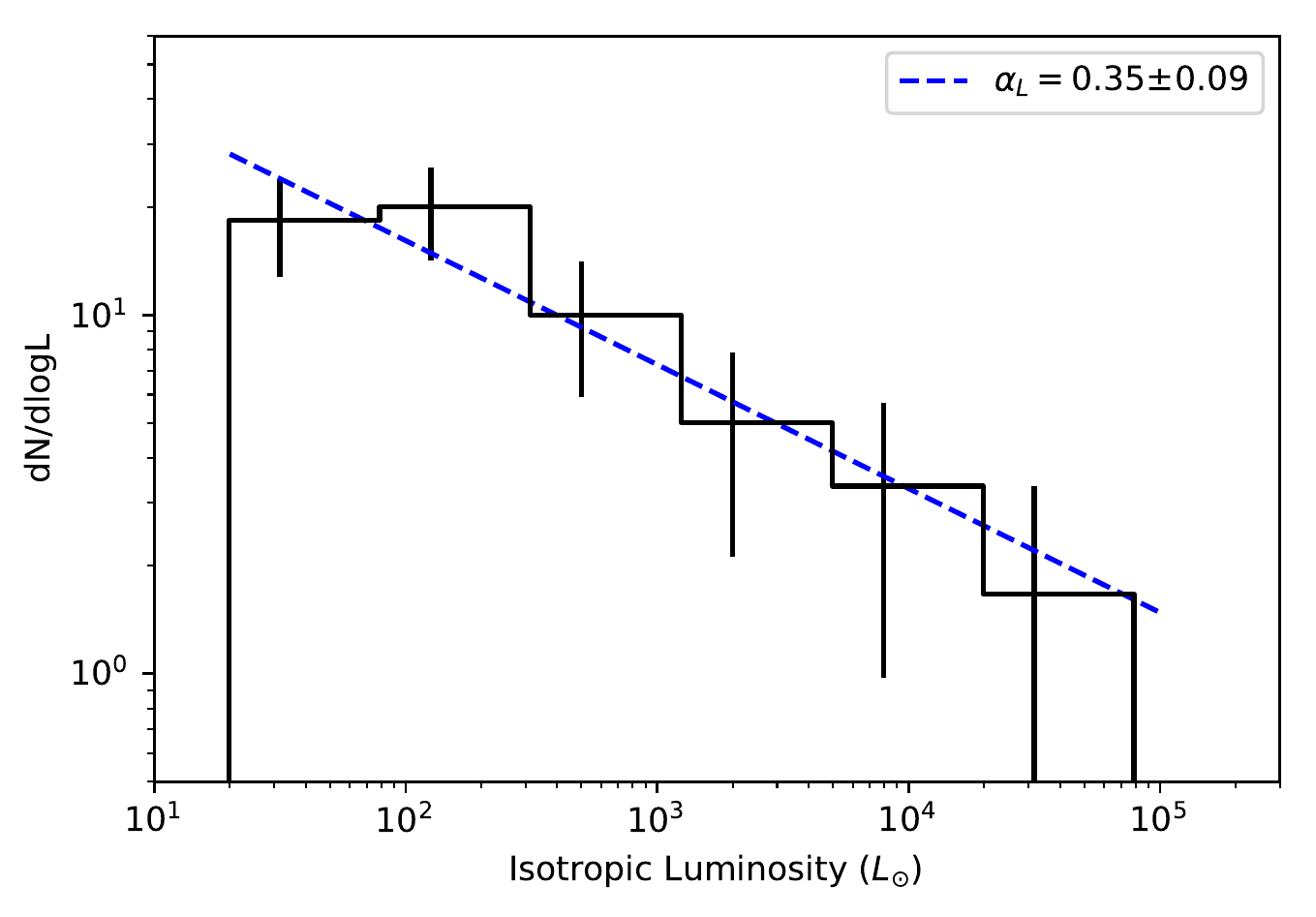}{0.45\textwidth}{}
          }
\vspace{-0.95cm}          
\gridline{\fig{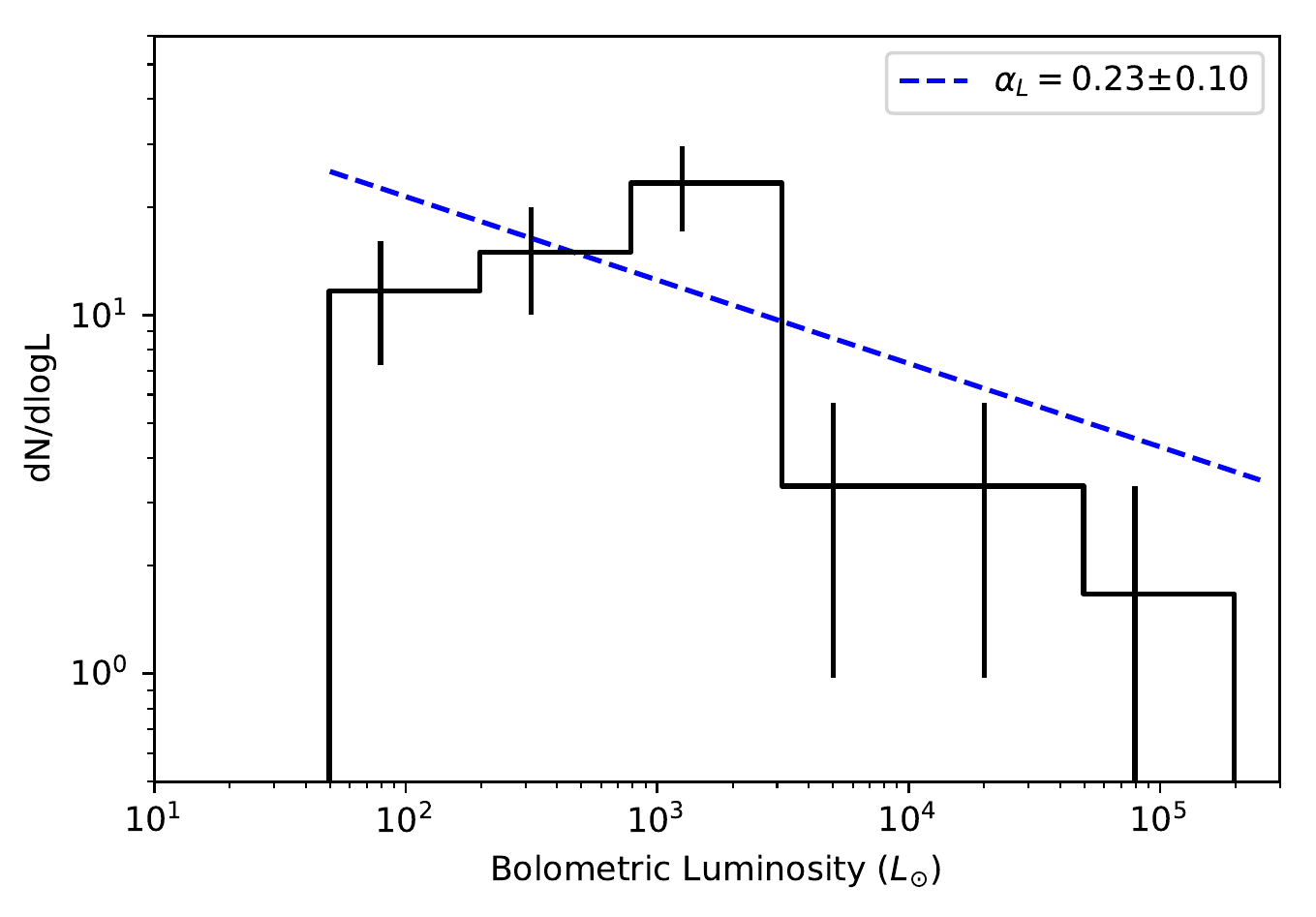}{0.45\textwidth}{}
	\hspace*{-1.6cm}
          \fig{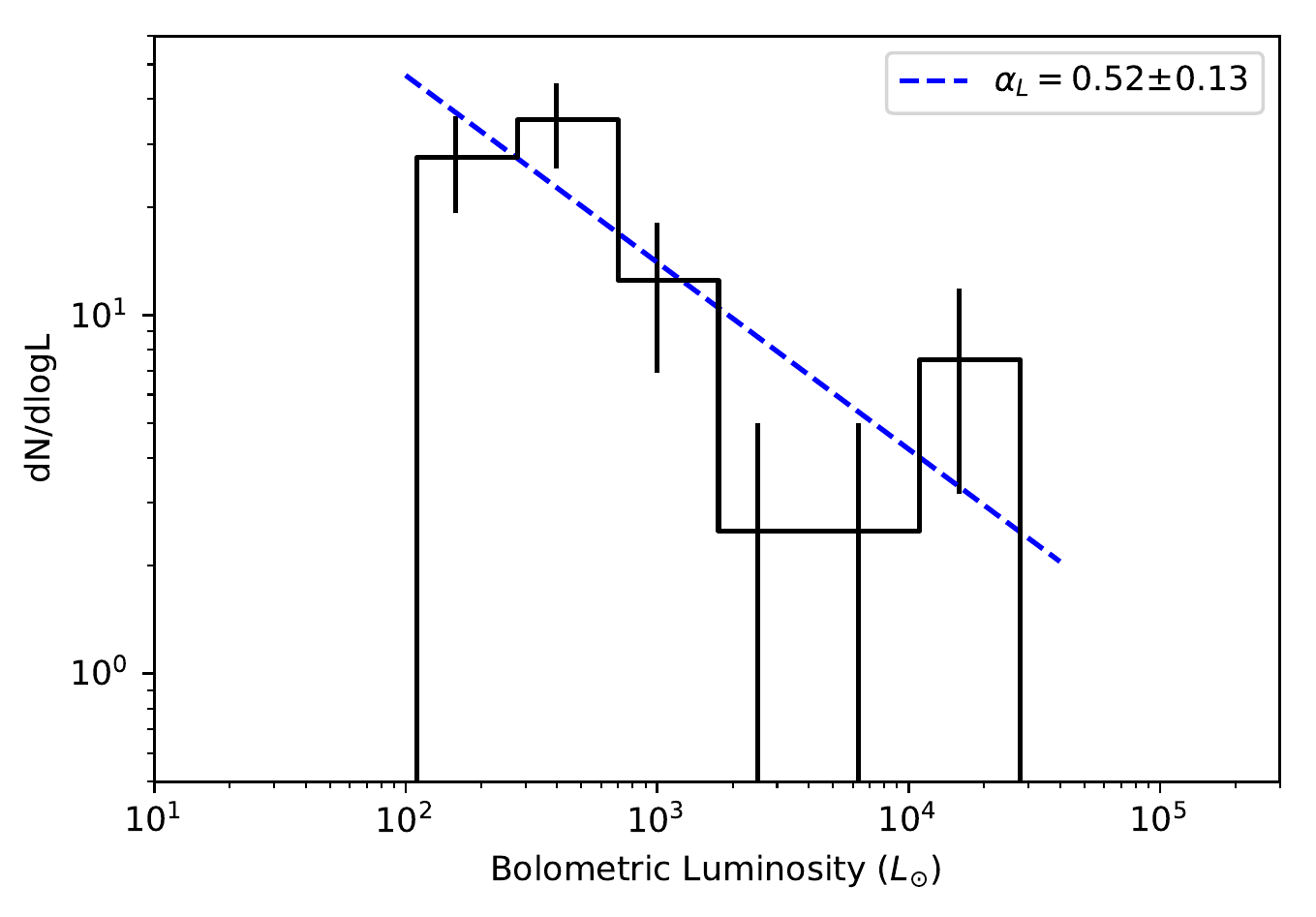}{0.45\textwidth}{}
          }
\caption{
{\it (a) Top Left:} Protostellar luminosity function, based on isotropic
luminosity $L_{\rm iso}$ values for the best model fits. The vertical
line in each bin shows Poisson sampling uncertainties. The dashed line
shows the best fit power law to the luminosity function of the form
$dN / d{\rm log}L\propto L^{-\alpha_L}$, with the result being
$\alpha_L=0.38\pm0.09$.
{\it (b) Top Right:} As (a), but now showing results for the average of ``good''
models, with $\alpha_L=0.35\pm0.09$.
{\it (c) Bottom Left:} As (a), but now showing the distribution of
model bolometric luminosities, based on best model fits, and with a
power law index of $\alpha_L=0.23\pm0.10$.
{\it (d) Bottom Right:} As (c),but now showing results for the average of ``good''
models, with $\alpha_L=0.52\pm0.13$.
}\label{fig:L}
\end{figure*}

We apply our fiducial analysis methods to all 35 identified sources in
the IRDC and list the derived protostellar properties (best fit model
and average of ``good'' models) in Table 3.

Figures~\ref{fig:L}a and b show the luminosity functions, based on
isotropic luminosity values, of the identified protostellar sources in
the IRDC. The luminosities range from $\sim3\times10^4\:L_\odot$,
i.e., for Cp23, down to $\sim30\:L_\odot$, i.e., sources similar to
Cp03. We fit a power law to the observed distribution of the form $dN
/ d{\rm log}L\propto L^{-\alpha_L}$, with the result being
$\alpha_L\simeq 0.35\pm0.09$ for the averages of ``good'' models. The
distributions are quite well fit by a single power law, with little
evidence for any break in the distribution, e.g., due to
incompleteness at low luminosities. Figures~\ref{fig:L}c and d show
the same information, but now for the distributions of bolometric
luminosities of the ZT models that are fit to the SEDs. The
distributions can still be fit with declining power laws, although
there now appears to be more deviation from simple, single power law
distributions. This is a reflection of the fact that the bolometric
model luminosity can be significantly different from the isotropic
luminosity, to both higher and lower values, due to beaming, i.e.,
``flashlight'', effects (see Zhang \& Tan 2018).

Our derived power law indices for the protostellar luminosity
functions are broadly consistent with those found by Eden et
al. (2018), who found the equivalent of $\alpha_L=0.26\pm0.05$ in W49A
and $\alpha_L=0.51\pm0.03$ in W51 (however, note there are significant
differences in their methods of source selection compared to our
70~$\rm \mu m$-based method).

\begin{figure*}
\gridline{\fig{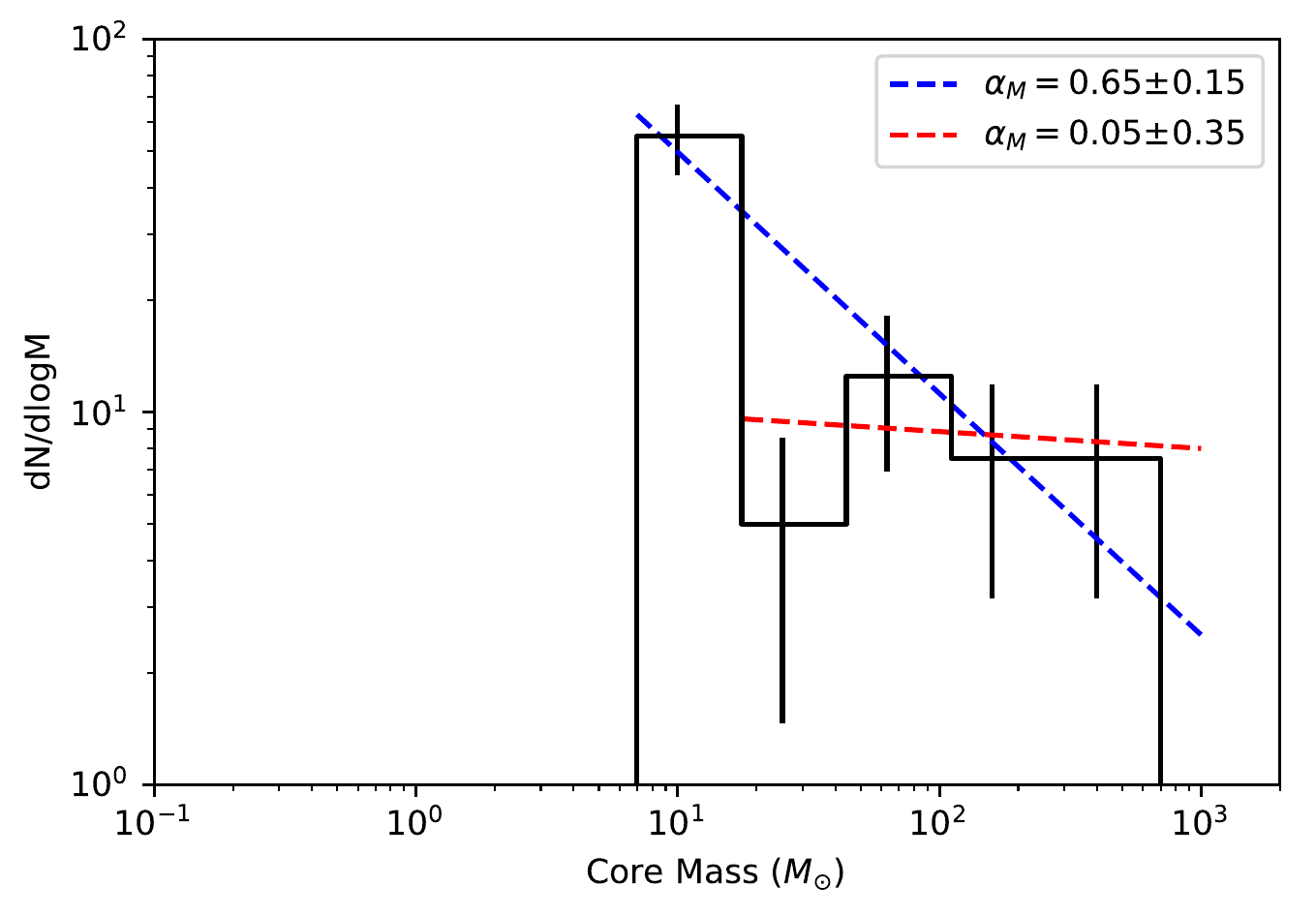}{0.45\textwidth}{}
	\hspace*{-1.6cm}
          \fig{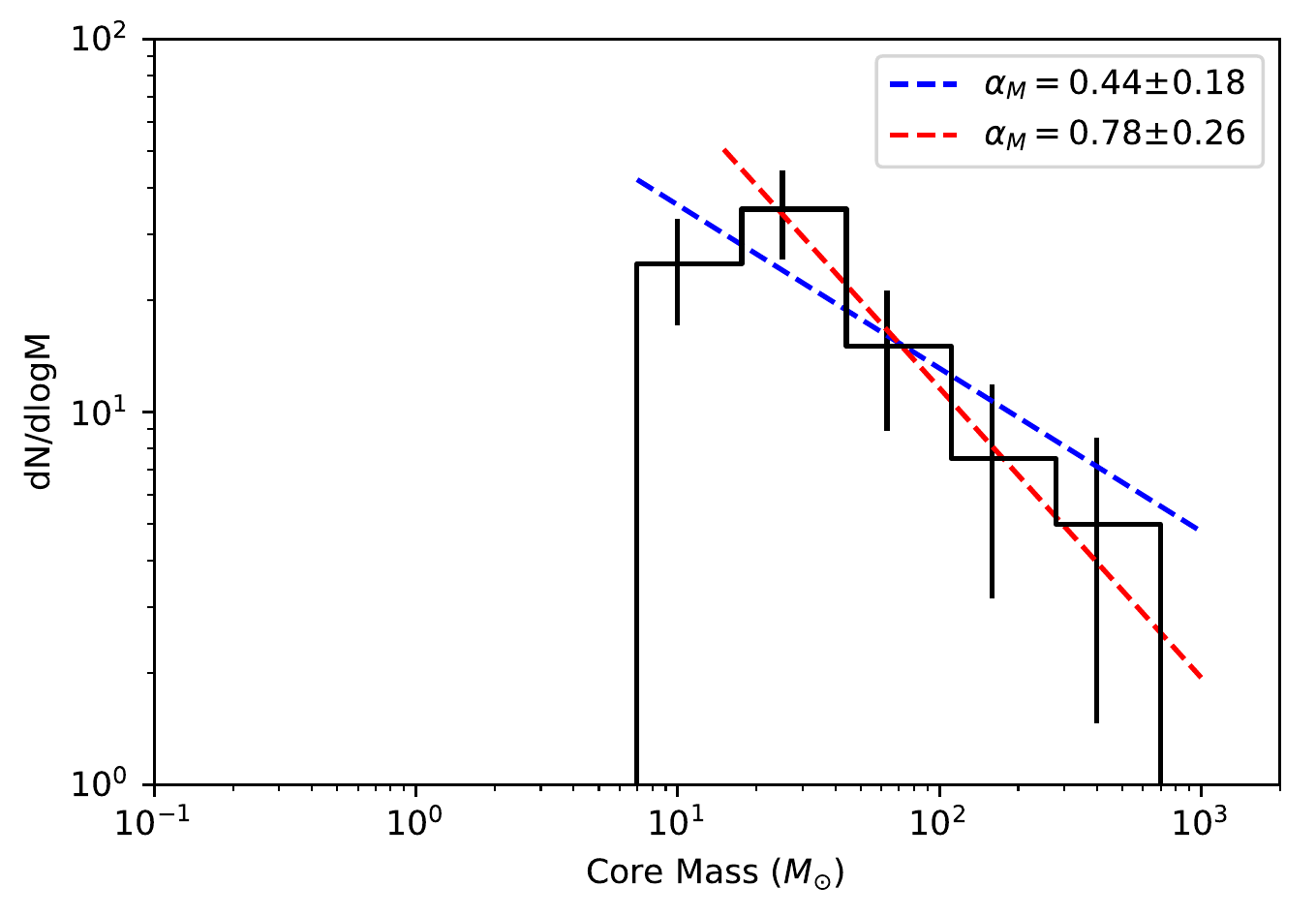}{0.45\textwidth}{}
          }
\vspace*{-0.9cm}          
\gridline{\fig{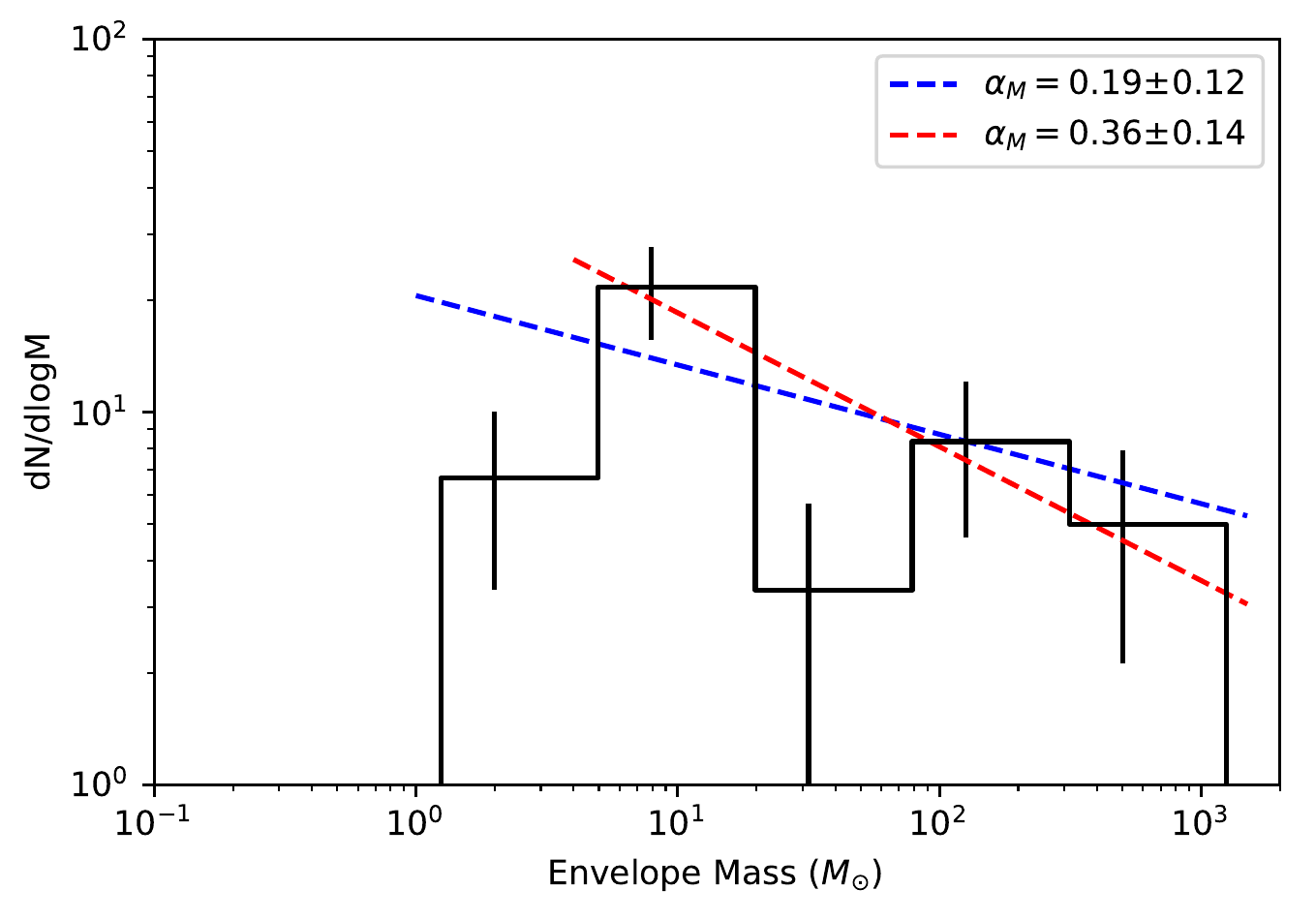}{0.45\textwidth}{}
	\hspace*{-1.6cm}
          \fig{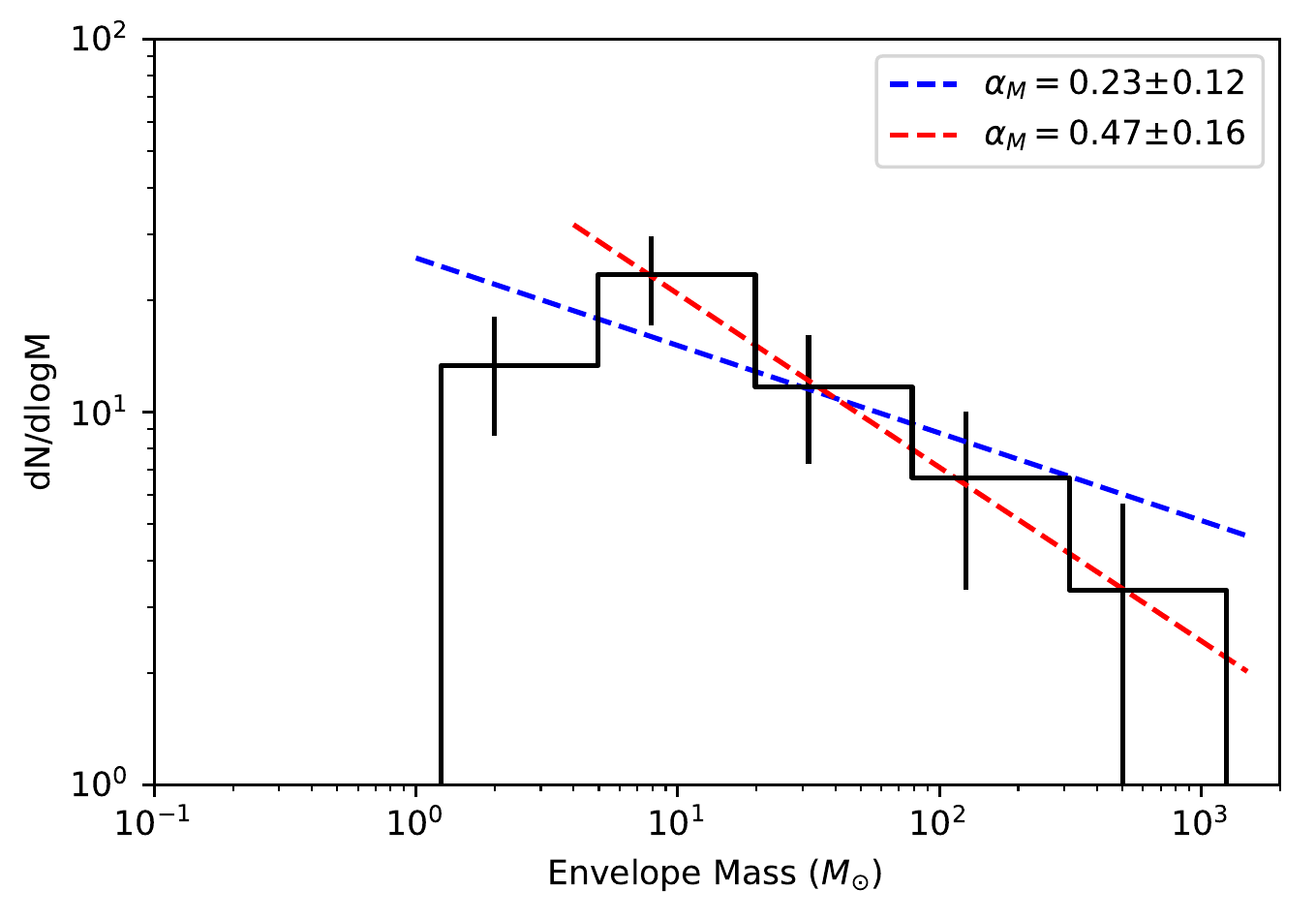}{0.45\textwidth}{}
          }    
\vspace*{-0.9cm} 
\gridline{\fig{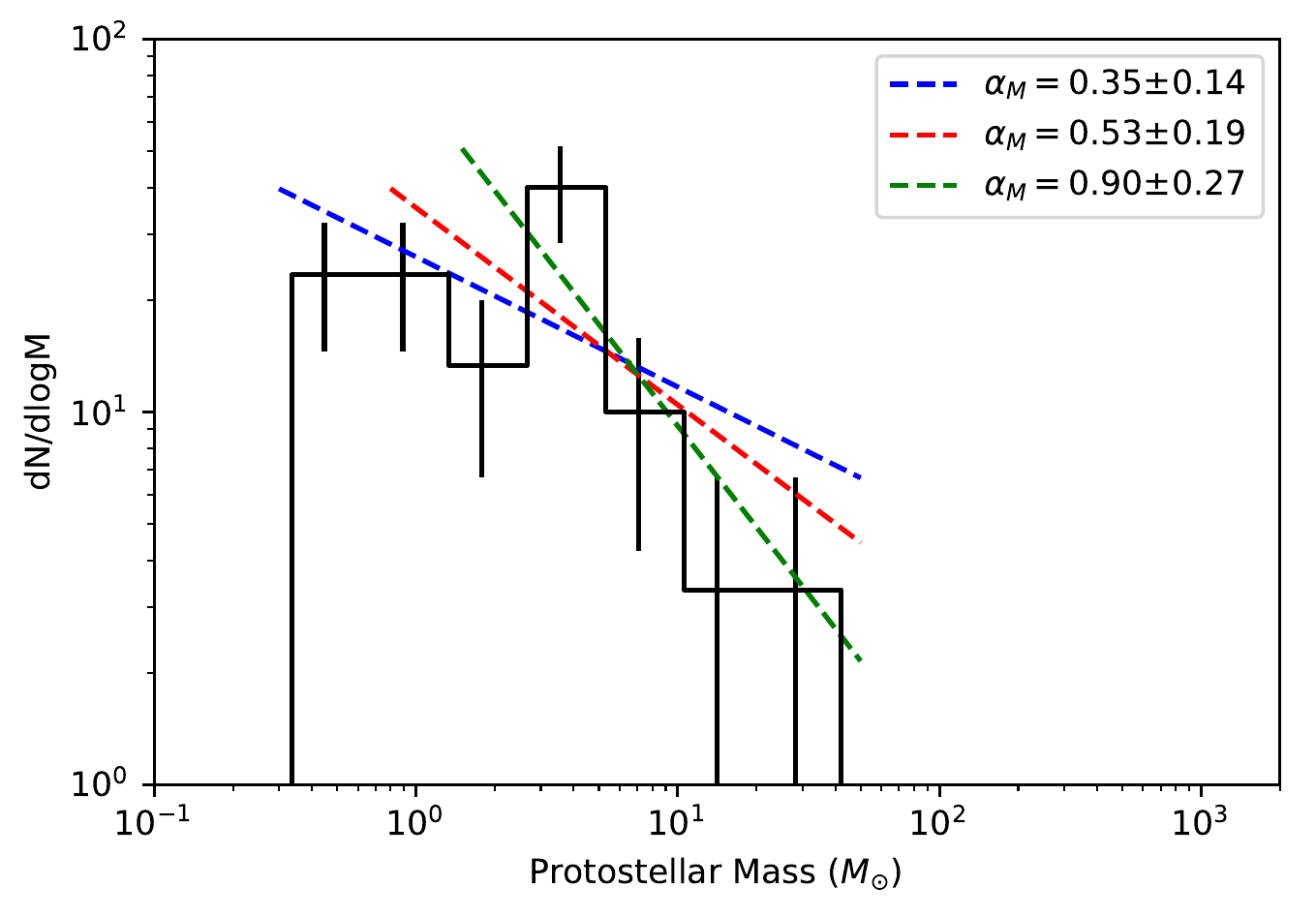}{0.45\textwidth}{}
	\hspace*{-1.6cm}
          \fig{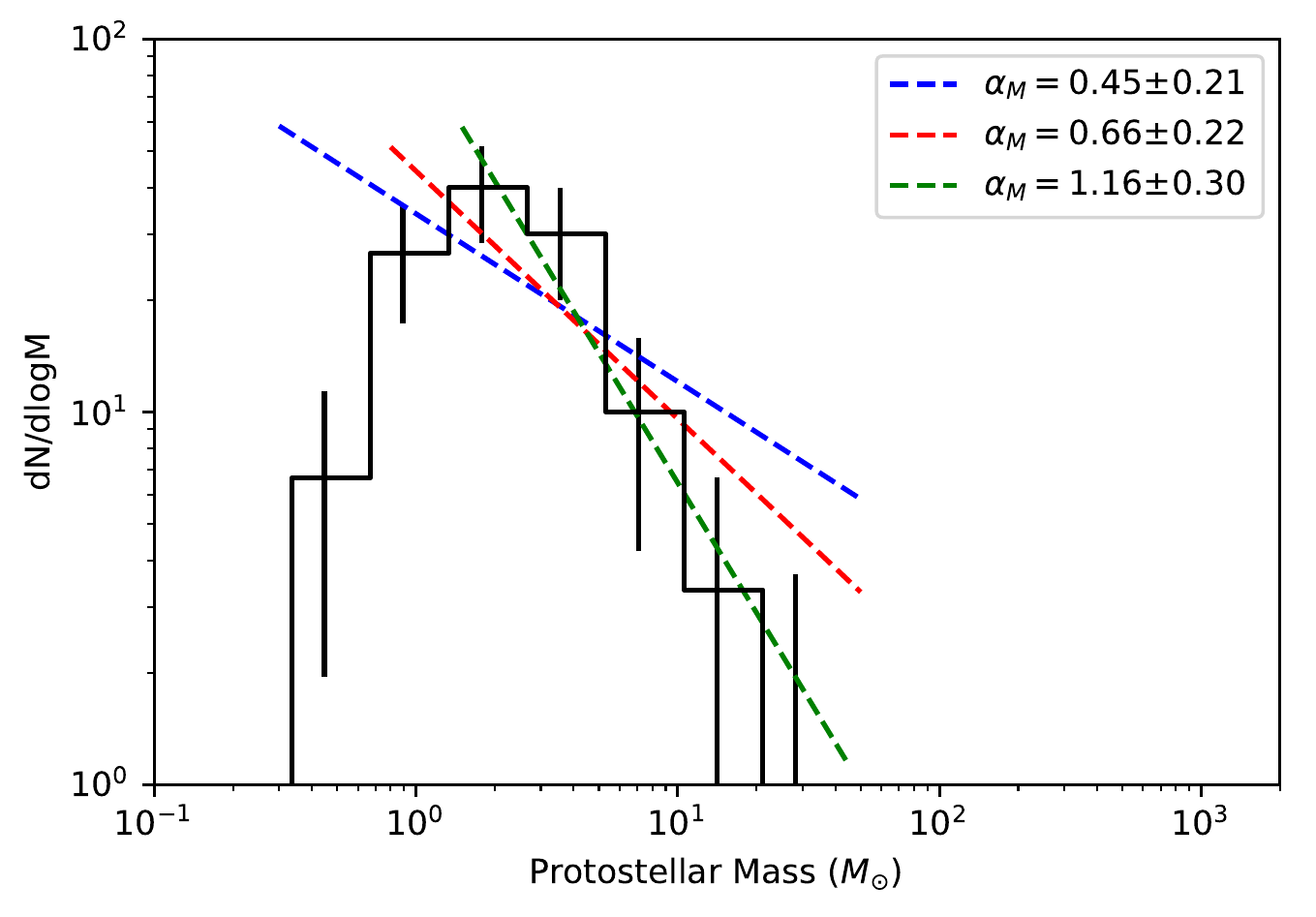}{0.45\textwidth}{}
          }       
\caption{
{\it (a) Top left:} Initial core mass function, as derived from best
fit models for each protostellar source. The vertical line in each bin
shows Poisson sampling uncertainties. The blue dashed line shows the best
fit power law to the full luminosity function of the form $dN / d{\rm
  log}M_c\propto M_c^{-\alpha_M}$, with the result being
$\alpha_M=0.65\pm0.15$. The red dashed line shows the same type of fit, but excluding the lowest mass bin.
{\it (b) Top right:} As (a), but now for averages of ``good'' model
fits, and with $\alpha_M=0.44\pm0.18$. We also show a fit that ignores
the lowest mass bin, which has $\alpha_M=0.78\pm0.26$.
{\it (c) Middle left:} As (a), but now for derived envelope masses of
the best fit models, and with $\alpha_M=0.19\pm0.12$.
{\it (d) Middle right:} As (c), but now for averages of ``good'' model
fits, and with $\alpha_M=0.23\pm0.12$. We also show a fit that ignores
the lowest mass bin, which has $\alpha_M=0.47\pm0.16$.
{\it (e) Bottom left:} As (a), but now for derived protostellar masses of
the best fit models, and with $\alpha_M=0.35\pm0.14$. The green dashed line shows a power law fit to the distribution excluding the two lowest mass bins.
{\it (f) Bottom right:} As (e), but now for averages of ``good'' model
fits, and with $\alpha_M=0.45\pm0.21$. We also show a fit that ignores
the lowest two mass bins, which has $\alpha_M=1.16\pm0.30$.
}\label{fig:M}
\end{figure*}

Figure~\ref{fig:M} shows the distributions of the protostellar
population in terms of their initial core masses (top row), current
envelope masses (middle row) and current protostellar masses (bottom
row), as derived from the model fitting, with best fit model results
shown in the left column and average of ``good'' model results shown
in the right. The minimum core mass in the model grid is
$10\:M_\odot$, which truncates the distribution at this point. This
may lead to a ``pile up'' in the distribution at the lower boundary,
which appears to be present in the distribution of best fit values of
$M_c$, but is not apparent for the average masses. We fit power laws
to the observed distributions of the form $dN / d{\rm log}M_c\propto
M_c^{-\alpha_M}$, with the result being $\alpha_M\simeq 0.44\pm0.18$
for the averages of ``good'' models. For comparison, the standard
Salpeter distribution of the stellar initial mass function (IMF) has
$\alpha_M=1.35$, with such values also being found for observed core
mass functions (CMFs) in some regions (e.g., Alves et al. 2007; Ohashi
et al. 2016; Cheng et al. 2018; Massi et al. 2019). We see that our
derived result for the initial core mass function in the IRDC
G028.37+00.07 is significantly shallower than the Salpeter
value. There have been claims of CMFs that are shallower, i.e., more
top heavy, than Salpeter in some star-forming regions: e.g., W43 by
Motte et al. (2018), who find $\alpha_M=0.90\pm0.06$; and dense IRDC
clumps by Liu et al. (2018), where a value of $\alpha_M=0.86\pm0.11$
has been reported. Still, our result of $\alpha_M\simeq 0.44\pm0.18$
is even flatter than these cases. It should be noted that it applies
over a higher mass range than has been probed by the Liu et al. (2018)
study. Also, our results here are based on indirect inference of the
initial protostellar core masses, while observational CMF studies,
including that of Liu et al., are based on direct observations of
cores.

To more closely connect with observational studies of the CMF, in
Figure~\ref{fig:M}c and d, we also show the core envelope mass
function of the identified protostars. Here the models now span to
masses below $10\:M_\odot$. For the average model results, the derived
power law index of the mass function is shallower than the initial core
mass function, which may be related to a larger mass range that is now
being probed and thus potentially great levels of incompleteness
affecting the lower mass regime. Still, fitting cores with envelope
masses $\gtrsim5\:M_\odot$, we still find a relatively shallow index
of $\alpha_M=0.47\pm0.16$.

Finally, the last two panels of Figure~\ref{fig:M} show the current
protostellar mass functions. The average results do not appear well
described by a single power law, perhaps because of incompleteness
and/or larger uncertainties at low masses. If we exclude the lowest
mass bin, we find $\alpha_M=0.66\pm0.22$, while excluding the two
lowest mass bins yields $\alpha_M=1.16\pm0.30$, which is consistent
with the Salpeter value of 1.35. However, it is already known that
deriving protostellar masses from SED fitting of massive protostellar
sources can suffer from a problem of high degeneracy (De Buizer et
al. 2017; Liu et al. 2019), and we see from the cases of Cp23, Cp15,
and Cp03 reported in Table 1 and Figure~\ref{fig:Cp23} and from the dispersions of
the protostellar masses listed in Table 3 that this problem persists
also for the lower luminosity sources that we fit here. Thus, caution
is needed when considering the value of $\alpha_M$ that is found from
this analysis, since it may be subject to change once more accurate
methods of estimating individual protostellar masses become available,
e.g., by dynamical means from study of their accretion disk gas
kinematics.

For the core masses, several points also need to be considered. Core
masses that are derived from SED model fitting results show quite a
wide dispersion in values among the best ten model fits. This is
expected since the models are mostly being constrained by the
luminosity of the source, much of which comes from warmer material
that does not dominate the core mass. Most of the core mass is at
larger distances from the source and thus at cooler temperatures and
so mostly affects the longer wavelength part of the SEDs. As shown in
Figure~\ref{fig:examples}, the model SEDs can often underpredict the
long wavelength part of the SED. This difficulty was already noted by
De Buizer et al. (2017) and ZT18. The cause may be due to imperfect
background subtraction at the longer wavelengths, especially when
model core radii are relatively small compared to source apertures.

\begin{figure*}
\gridline{\fig{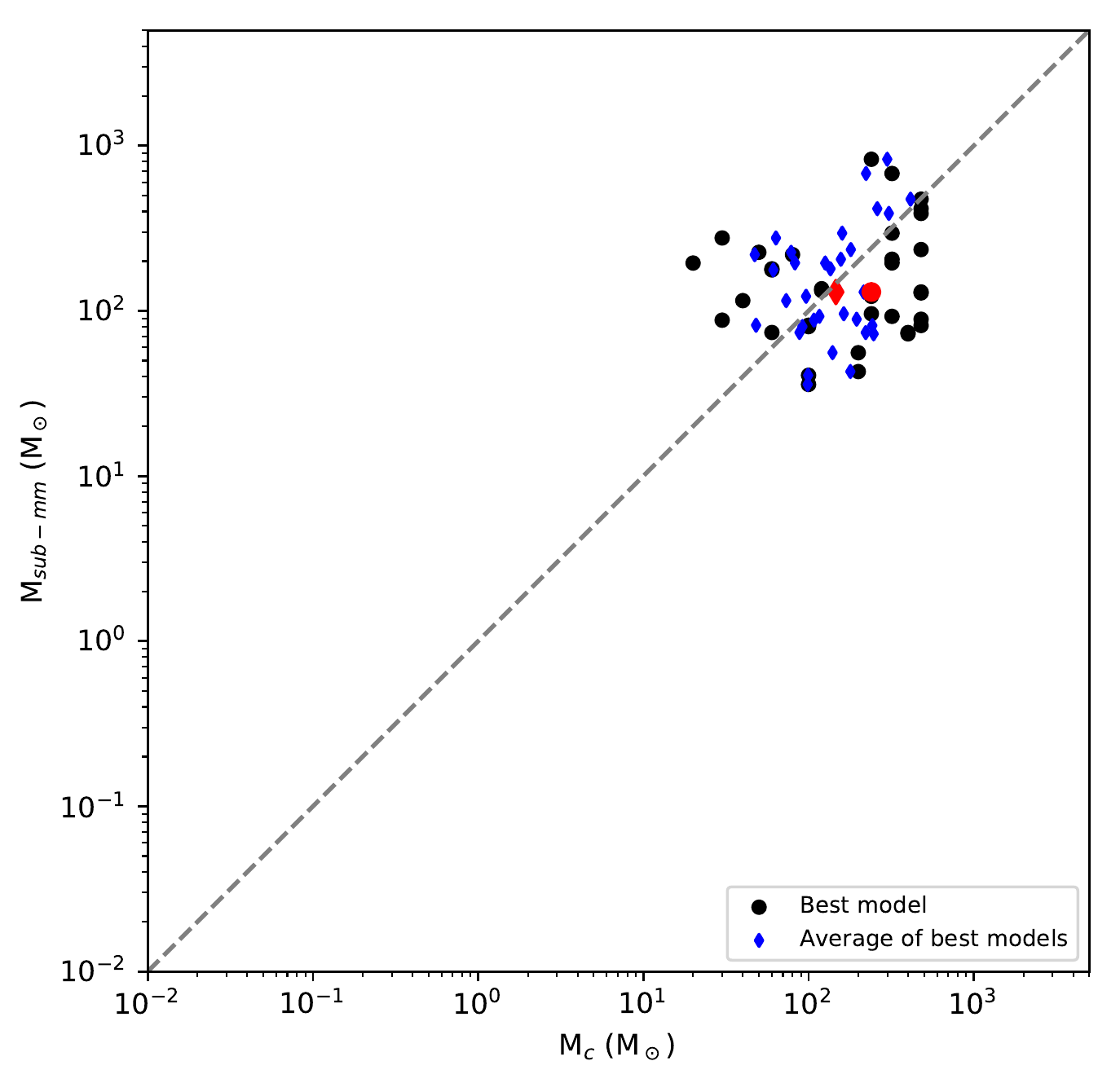}{0.45\textwidth}{}
	\hspace*{-1.6cm}
          \fig{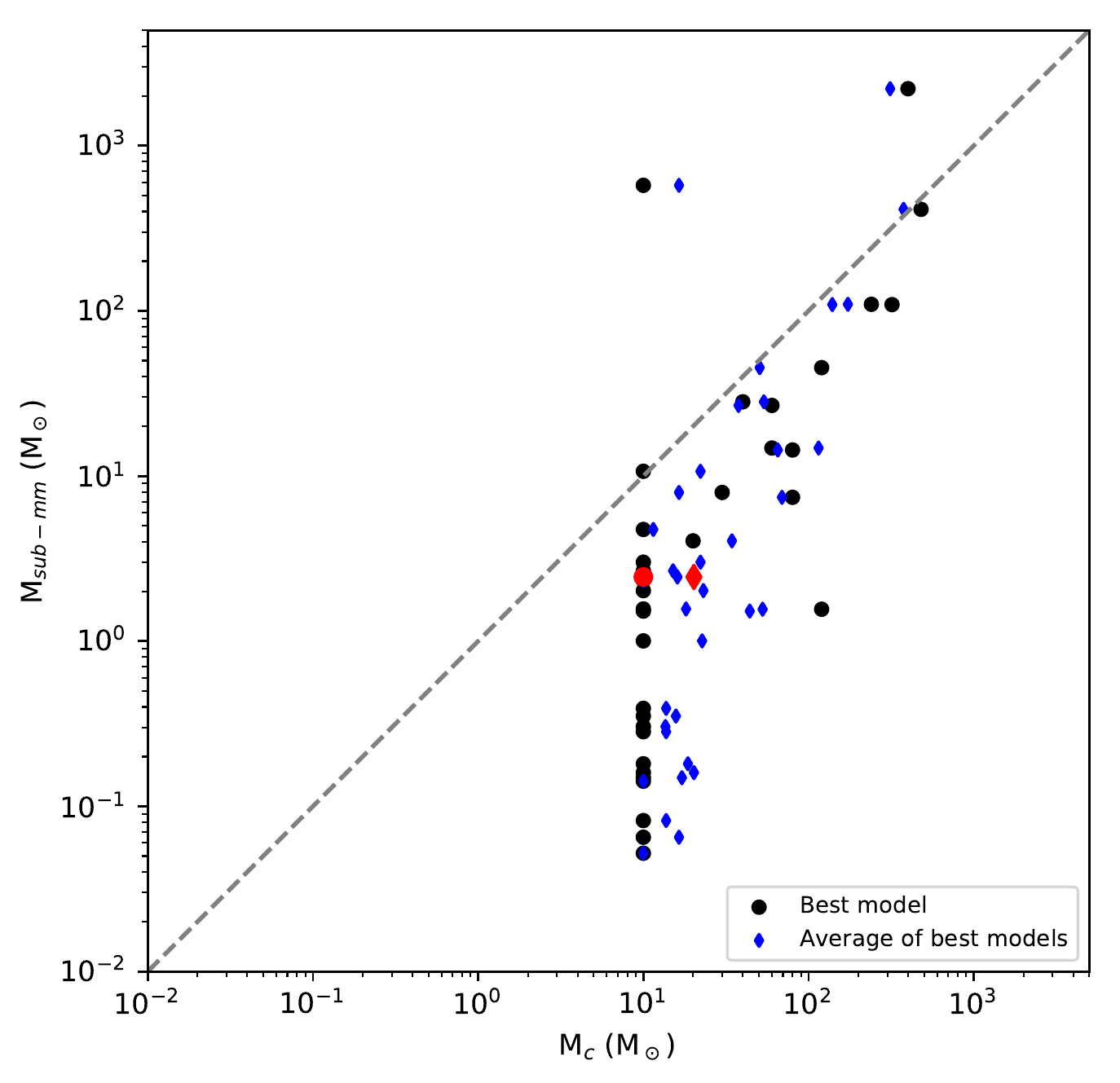}{0.45\textwidth}{}
          }
\vspace*{-0.9cm}           
\gridline{\fig{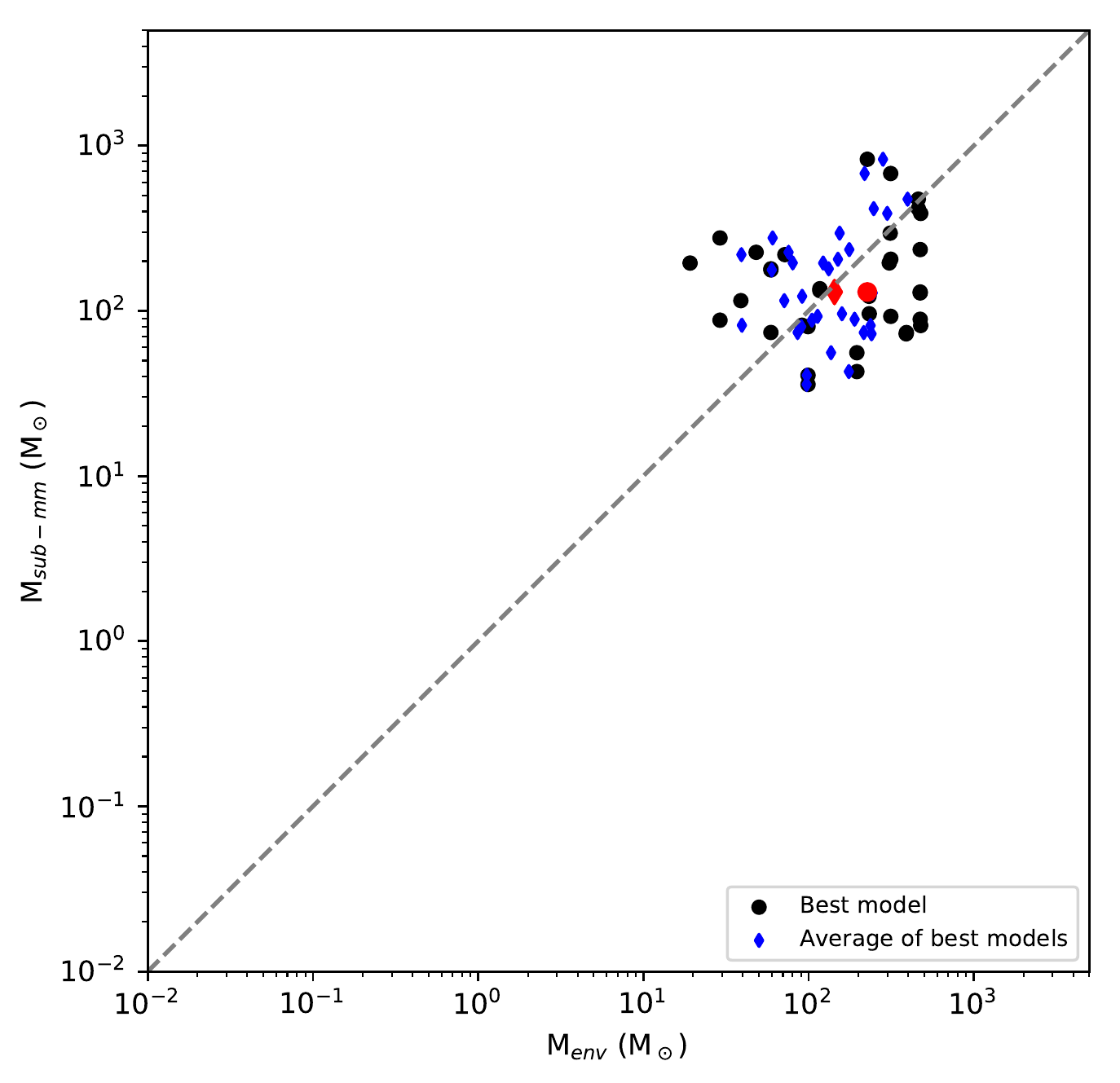}{0.45\textwidth}{}
	\hspace*{-1.6cm}
          \fig{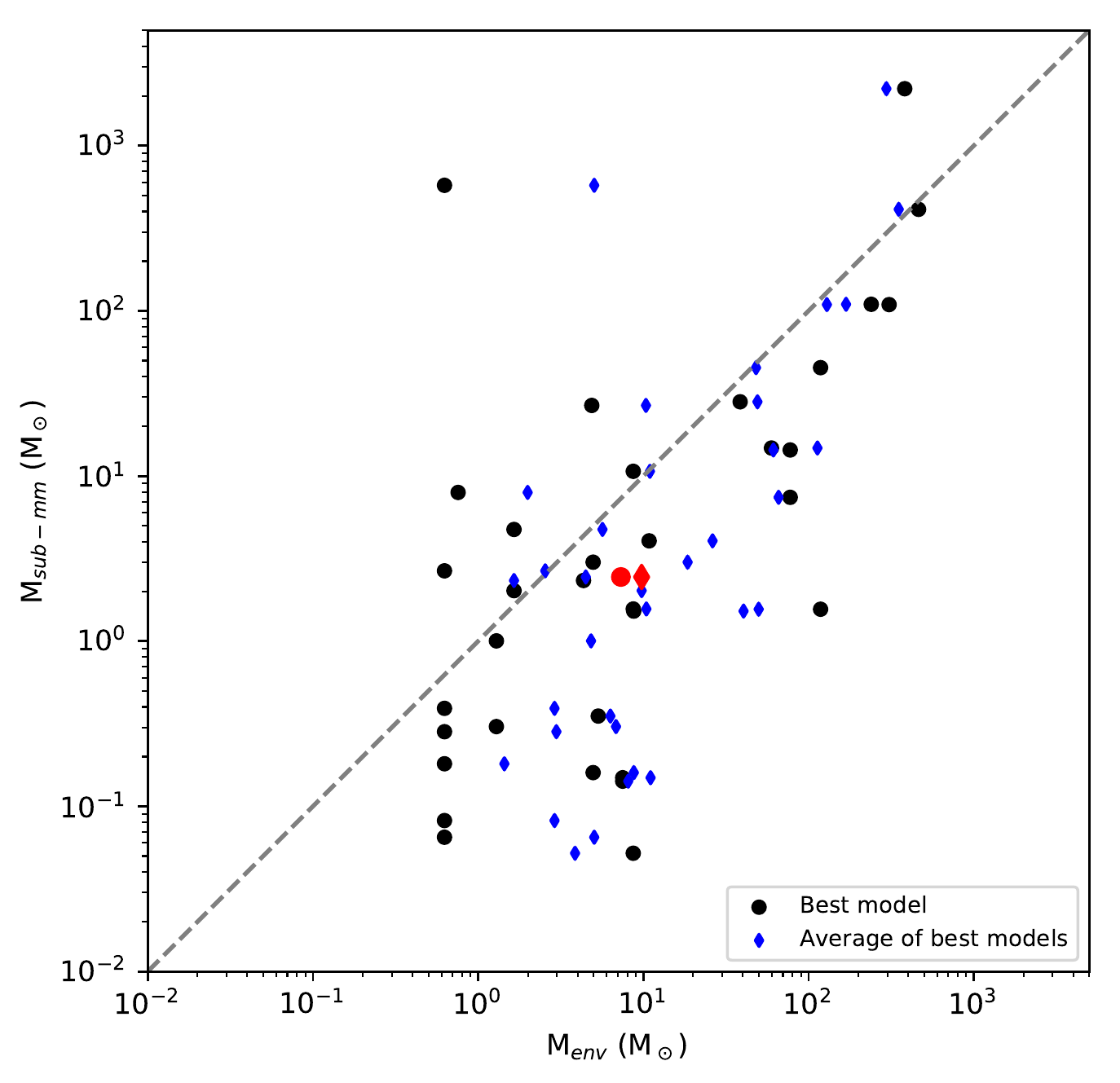}{0.45\textwidth}{}
          }
\caption{
Comparison of single temperature greybody derived masses, $M_{\rm
  submm}$, with $M_c$ (top row) and $M_{\rm env}$ (bottom
row). Results before background subtraction are shown in the left
column. Results after background subtraction, i.e., the fiducial case,
are shown in the right column. In each panel, the best model (black
circle) and average of ten best models (blue diamond) are shown for
each source. The red circles and diamonds show the median values of
these metrics, respectively. The one-to-one line is also displayed for
reference.  The large effect of background subtraction on $M_{\rm
  submm}$ is apparent. Also visible is the pile-up of $M_c$ values at
the minimum of $10\:M_\odot$.
}
\label{fig:masscomparison}
\end{figure*}

A more direct mass estimate from a given SED can be made by carrying
out a single temperature greybody fit to just the longer wavelength
component of the SED, i.e., from 160 to 500~$\rm \mu m$, following the
methods of, e.g., Lim et al. (2016). Comparisons of these mass
estimates, i.e., $M_{\rm submm}$, with those resulting from the ZT
model values for $M_c$ and $M_{\rm env}$ are shown in
Figure~\ref{fig:masscomparison}. This figure shows the large effect of
background subtraction on $M_{\rm submm}$. Also visible is the pile-up
of $M_c$ values at the minimum value of $10\:M_\odot$, which is simply
an artifact of a limitation of the ZT model grid. The best agreement
is expected between background subtracted values of $M_{\rm submm}$
and $M_{\rm env}$ and indeed this is apparent in the lower right panel
of Figure~\ref{fig:masscomparison}. Still, this comparison shows there
is significant scatter in the correlation and with a modest systematic
offset of $M_{\rm submm}$ being lower than $M_{\rm env}$ by a factor
of a few on average.

\begin{figure*}
\includegraphics[scale=0.85]{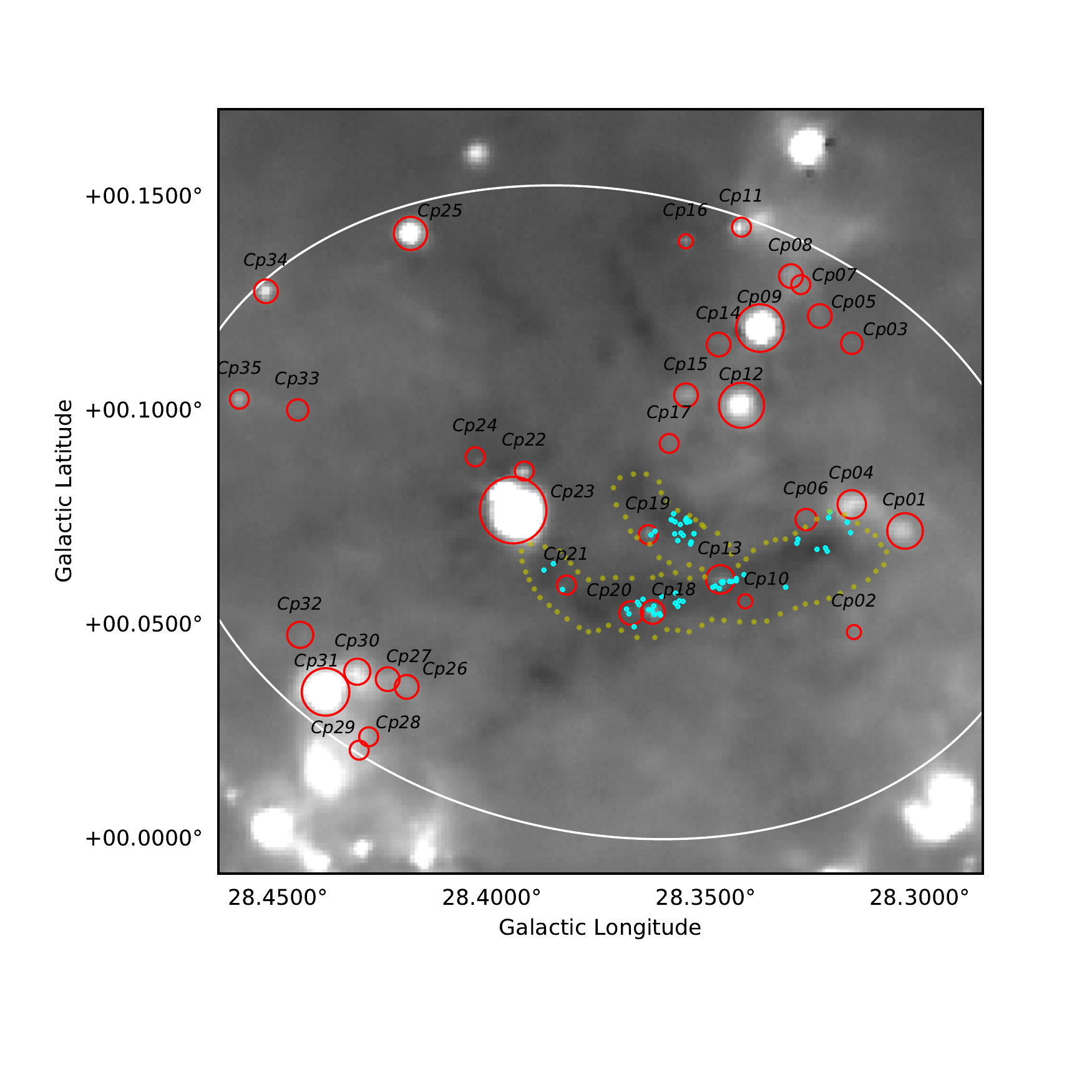}
\vspace{-1.6cm}
\caption{Comparison with Kong et al. 2019 outflows: Kong sources shown in cyan overlaid on 70$\mu$m image of Cloud C and region matching the ALMA primary-beam response at 30$\%$ (outer contour of Figure 1 in Kong et al. 2019) shown here in yellow.}
\label{fig:kongcomparison}
\end{figure*}

\subsection{Comparison with ALMA outflow observations}\label{S:alma}

A recent study of molecular outflows in the central region of Cloud C
was performed by Kong et al. (2019). A comparison of our $70\:{\rm \mu
  m}$ {\it Herschel}-defined sources with the {\it ALMA}-detected
outflow-driving sources is shown in Figure~\ref{fig:kongcomparison}.

We see that in some of the {\it Herschel}-identified sources, i.e.,
Cp13, Cp18, and Cp20, there are actually several different {\it
  ALMA}-identified protostars present. There are two {\it
  ALMA}-protostars in Cp19 and only one in Cp21. Of the six {\it
  Herschel}-identified protostars that are fully covered by the {\it
  ALMA} data, only Cp10 does not have an {\it ALMA}-identified
protostar. At the same time, the {\it Herschel}-identification method
also misses a significant amount of protostellar activity: about 2/3
of the {\it ALMA} sources are not associated with a {\it
  Herschel}-identified source.
It should be noted that this comparison has been done for just a small
number of {\it Herschel}-identified sources and in a relatively small
part of the IRDC. Also it is a particularly MIR and FIR (including
$70\:{\rm \mu m}$) dark region, with the {\it Herschel} sources being
of relatively low luminosity.

These results indicate that the association of a {\it
  Herschel}-identified protostar, i.e., based on a relatively low
angular resolution imaging of dust continuum emission, can often be
problematic, at least for sources at $\sim$5~kpc distances, like IRDC
G028.37+00.07. This is mitigated somewhat when there are just a few
sources in the aperture and one of them is clearly the dominant
source, which may be the case in Cp19 based on the intensity of the
outflows.

Ultimately, more accurate protostellar SED characterisation will
require higher angular resolution observations that cover the peak
wavelength range of the SED. Still, with these caveats in mind we
proceed to derive the overall star formation activity that is implied
by the population of {\it Herschel}-identified sources.

\newpage
\subsection{Star Formation Rate and Efficiency}

The star formation rate (SFR), including as a star formation
efficiency (SFE) per free-fall time ($\epsilon_{\rm ff}$), is
important to quantify, e.g., as a constraint on theoretical models of
star cluster formation.

The total mass of Cloud C is measured from extinction mapping to be
68,300 $M_\odot$ as given by BTK14, with 50 percent overall
uncertainty (dominated by systematics). Using a different method based
on sub-mm dust emission, Lim et al. (2016) estimated the mass to be
72,000~$M_\odot$. We will adopt a total mass of Cloud C in the defined
ellipse region of 70,000~$M_\odot$.

Summing all of the masses resulting from the ZT models (Table 3), the
total mass of the sources in the cloud is 1642~$M_\odot$ of total
initial protostellar core mass for best-fitting models and
1740~$M_\odot$ of total initial protostellar core mass for average of
good models. About 50\% of this mass is expected to eventually go into
stars (Tanaka et al. 2017). However, the results of \S{\ref{S:alma}}
indicate a completeness correction factor with respect to the {\it
  ALMA}-detected population of about three needs to be accounted for
(or potentially a smaller factor if the missing sources tend to be
lower-mass protostars). However, the {\it ALMA}-identified population
is also likely to be incomplete at some level. Thus by this method we
estimate that the protostellar population that is forming in the IRDC
will produce a mass of stars of about 2,000~$M_\odot$. This represents
2.9\% of the total IRDC mass.

However, since the ZT model grid is designed for higher-mass
protostellar cores (i.e., $>10\:M_\odot$), it is possible that the
above results are biased towards too high core masses on average. As
an alternative method, we can consider the current protostellar masses
implied by the models.  If we sum the current protostellar masses then
we obtain 132.5~$M_\odot$ for best models and 86.8~$M_\odot$ for
average of good models, i.e., $\sim 100\:M_\odot$. Then with a factor
of 2 correction between current and final protostellar mass and a
factor of 3 correction due to incompleteness, we would estimate a
total mass of stars that will form of $\sim600\:M_\odot$, i.e., 0.86\%
of the total IRDC mass.

In the context of the Turbulent Core Model of McKee \& Tan (2003
[MT03]), the protostellar formation time is approximately 37\% of the
mean free-fall time of the clump, $\bar{t}_{\rm ff,cl}$ (eq. 37 of
MT03), based on the mass of a $10\:M_\odot$ star forming from a
$20\:M_\odot$ core (the formation time scales weakly as
$m^{1/4}$). Thus, assuming the protostellar population we have sampled
traces the activity of protostars forming in the last 40\% of $t_{\rm
  ff,cl}$, we estimate a star formation efficiency per free-fall time
in the IRDC of between 2.1\% and 7.3\%, depending on the
above methods of mass estimation.

We consider that the lower estimate here is more reliable, since it is
tied more closely to the protostellar luminosities and avoids the
expected bias of too high initial core masses that will be found from
using the ZT grid. If we have included some already formed stars that
are present in the IRDC and that are simply heating surrounding local
IRDC material, then we will have overestimated the SFR. On the
other hand, the uncertain ALMA incompleteness factor would boost the
estimate. Overall, we consider that the data support an estimate of
$\epsilon_{\rm ff}\sim 0.03$ in IRDC C.

If the IRDC is to form a bound cluster, which may be a reasonable
expectation since it is one of the most extreme IRDCs and does appear
to be gravitationally bound and in approximate virial equilibrium at
the moment (BTK14; Hernandez \& Tan 2015), then an overall star
formation efficiency (SFE) of $\gtrsim30\%$ is likely to be needed. At
the current SFR this would then take $\sim10$ free-fall times to be
achieved. 

The absolute value of the free-fall time is measured as $1.3 \times
10^{6}$ years, using the equation 
$t_{ff}=[3\pi/32G\rho]^{1/2}$
and the properties measured by BTK14. Then the total star formation rate implied by $\epsilon_{\rm
  ff}=0.03$ is $1.6\times10^{-3}\:M_\odot\:{\rm yr}^{-1}$. Thus, age
spreads of at least 1~Myr are expected, even in fastest formation
models, but closer to 10~Myr if a bound cluster is to form with our
above estimate of $\epsilon_{\rm ff}=0.03$. However, the age spread
could be reduced if the protocluster clump evolves to a denser state
that has a shorter local free-fall time, which would then lead to an
increasing absolute SFR (see, e.g., Palla \& Stahler 2000).

\subsection{Spatial Distribution of Protostars}\label{S:spatial}

The initial spatial distribution of stars within forming clusters,
including degree of substructure, central concentration and primordial
mass segregation, is of interest to help constrain theoretical models
of both massive star formation (i.e., are special conditions needed
for massive star formation) and for star cluster formation. However,
it is in general difficult to infer these properties from observations
of already formed stars, because signatures of the initial conditions
are erased by dynamical evolution. As far as we are aware, there are
no measurements yet of these properties based on protostellar
populations in massive ($>10^4\:M_\odot$) protoclusters.

One widely used parameter to measure substructure is the $\cal Q$
parameter (Cartwright \& Whitworth 2004), which is the ratio between
the mean length of the edges of the minimum spanning tree (MST) of the
cluster, $\bar{m}$, and the mean separation between stars in the
cluster, $\bar{s}$. This parameter has the ability to distinguish
between a substructured regime and a radially concentrated regime. A
value ${\cal Q} < 0.785$ means the cluster is relatively substructured
with a lower value corresponding to more clumpiness. In contrast,
${\cal Q} > 0.785$ means the cluster has an overall radial
structure/concentration, with a higher $\cal{Q}$ value indicating that
it is more concentrated in the center.

We measure a value of ${\cal Q} = 0.667$ for the 35 protostellar
sources of Cloud C. This value classifies the cluster as
``substructured'', i.e., comparable to a three dimensional
distribution with a fractal dimension $D\sim 2$, i.e., considerably
substructured and not centrally concentrated (see Cartwright \&
Whithworth 2004).  As stated in \S2, the total number of sources
identified by the Hi-GAL catalog was 40, but then was reduced to 35
due to crowding and lack of resolution for our aperture photometry
analysis. However, if we consider the case of all 40 sources, the
${\cal Q}$ parameter changes only modestly to 0.640, resulting in the
same basic classification for the degree of substructure.

This value of $Q\simeq 0.67$ falls within the middle of the range of
values reported by Cartwright \& Whitworth (2004) for lower-mass
clusters, i.e., Taurus (0.47), IC2391 (0.66), Chameleon (0.67), $\rho$
Ophiuchus (0.85) and IC348 (0.98). This may indicate that whatever
process controls the initial distribution of protostars does not vary
significantly across the star-forming clump mass spectrum.

Our observed value of $Q$ can be compared to that seen in numerical
simulations of cluster formation. For example, Wu et al. (2017)
studied cluster formation from colliding and non-colliding GMCs. In
the colliding case, the simulations showed values of $Q$ that
fluctuated in the range from $\sim0.3$ to $\sim1.5$, but often
settling at values near 0.6 (Fig. 9 of Wu et al. 2017). However, in
the non-colliding models the values of $Q$ were typically much smaller
at $\sim0.2$. It should be noted that these simulations did not
include feedback from the forming stars and were based on particular
initial conditions of quite idealized GMCs. Nevertheless, in the
context of these models, the colliding cases were able to form more
concentrated clusters that are closer to the observed systems,
including our result for the massive protocluster forming in IRDC
G028.37+00.07.

Considering the degree of mass segregation of the protostellar
population,
the simplest approach is to examine where the most massive stars are
with respect to a defined center of the cluster. In the case of IRDC
G028.37+00.07, while there is a center of the elliptical region that
has been used for defining the IRDC, it must be noted that this
definition, originally based on low resolution {\it MSX} images of the
Galactic plane (Simon et al. 2006), is somewhat arbitrary. Still, we
consider the three sources with the highest current protostellar mass
estimates (based on averages of ``good'' models; see Table~3), which
are: Cp09 with $m_*=10.5\:M_\odot$; Cp12 with $m_*=9.8\:M_\odot$; and
Cp23 with $m_*=8.0\:M_\odot$. The source Cp09 is located near the
ellipctical boundary of the IRDC; Cp12 is at an intermediate distance
from the center; while Cp23 is relatively close to the center of the
cloud (at least in projection). These results do not support there
being any strong preference for massive stars to form in the center of
the protocluster, at least as defined by the Simon et al. (2006)
ellipse.

Given the difficulty of defining a cluster center, an alternative
approach to study mass segregation is to think of it as the tendency
of massive stars to stay near other massive stars. This definition
does not need a defined center and so is better suited to deal with a
substructured protocluster. A popular method to measure mass
segregation in this way is the $\Lambda_{\rm MSR}$ parameter, which
also uses the MST (Allison et al. 2009).
This parameter compares the total length of the MST of the $N$ most
massive stars in the cluster to the length of the MST of $N$ stars in
the cluster picked randomly. To reduce variation caused by the random
selection of stars, this parameter is measured multiple times, which
also allows an estimate of its uncertainty (see Allison et al. 2009
for details). Then, a mass-segregated cluster has values of
$\Lambda_{\rm MSR} > 1$, a cluster with no mass segregation has
$\Lambda_{\rm MSR} = 1$ and a cluster with inverse mass segregation
has $\Lambda_{\rm MSR} < 1$, i.e. having the $N$ most massive stars
more separated in comparison with the average star. However, while
this method does not require a defined cluster center, it does require
defining a population of sources, which in our case has been done with
the condition that they are inside the already defined IRDC
boundary. We will return to this point below.

\begin{figure}[h]
\includegraphics[scale=0.6]{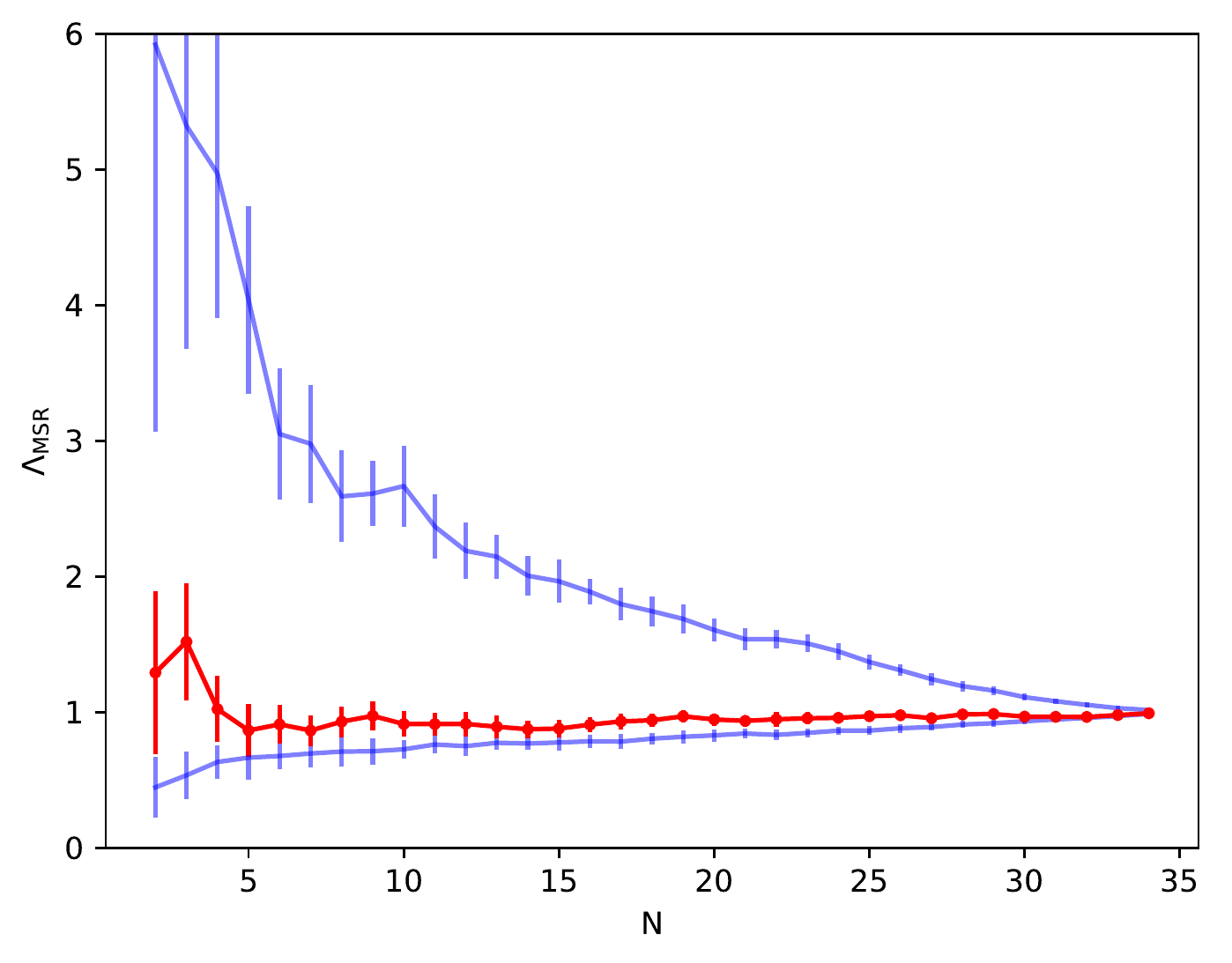}
\vspace{-0.5cm}
\caption{
Mass segregation parameter, $\Lambda_{\rm MSR}$, as a function of
number of stars, $N$, used to define the high-mass sample for the
protostars in IRDC G028.37+00.07 (red points and line). Masses are
based on averages of ``good'' protostellar masses for the sources. The
vertical line attached to each point shows the dispersion in results
given that the method involves random sampling of $N$ sources from the
total population (see \S{\ref{S:spatial}}). The blue points and lines show the most
extreme values of $\Lambda_{\rm MSR}$ that are possible given freedom
to reassign masses among the protostars at their observed locations.
}
\label{fig:msr}
\end{figure}

For calculation of $\Lambda_{\rm MSR}$, we focus on the estimates of
current protostellar masses based on average of ``good'' models (see
above). Figure~\ref{fig:msr} shows $\Lambda_{\rm MSR}$ as a function
of the number, $N$, of the most massive sources used to define the
high-mass sample. The figure also shows the maximum and minimum
possible values of $\Lambda_{\rm MSR}$ based on the locations of the
protostars, but with the freedom to reassign the masses to achieve
these extreme values. We see that $\Lambda_{\rm MSR}$ has values close
to one for $N\geq 4$, and with a modest enhancement of $\Lambda_{\rm
  MSR}\simeq1.4$ when $N\leq 3$. This is tentative evidence for a
signature of mass segregation at these numerical scales. However,
given the size of the uncertainties, this cannot be regarded as strong
evidence for a signature of primordial mass segregation (i.e.,
enhanced clustering) of the massive protostars in the IRDC. Still,
these results provide basic constraints with which to test theoretical
and numerical models of star cluster formation.

We have also investigated the sensitivity of these results to the
choice of IRDC boundary location. In particular we examine if the
results change if we exclude the seven sources in the SE region of the
IRDC, which are quite well separated from the main IRDC features
  in a relatively IR bright region and may be part of another
grouping of protostars that is seen just outside the IRDC boundary. We
find that the dependence of $\Lambda_{\rm MSR}$ versus $N$ shows very
similar behavior when we repeat the analysis on the remaining
sources. Thus the grouping of seven sources near the SE boundary was
not significantly affecting these clustering results. However, of
course the results for $\Lambda_{\rm MSR}$ versus $N$ could
change if the actual physical ``protocluster'' had a larger extent
that included a population of massive protostars that dominated over
those in the local IRDC region that we have focussed on. However, in
this case we would still conclude that any clustering of such
protostars is not especially concentrated toward the IRDC and that the
protostars in the IRDC region are themselves not especially clustered
or centrally concentrated.

\begin{figure*}
\includegraphics[scale=0.8]{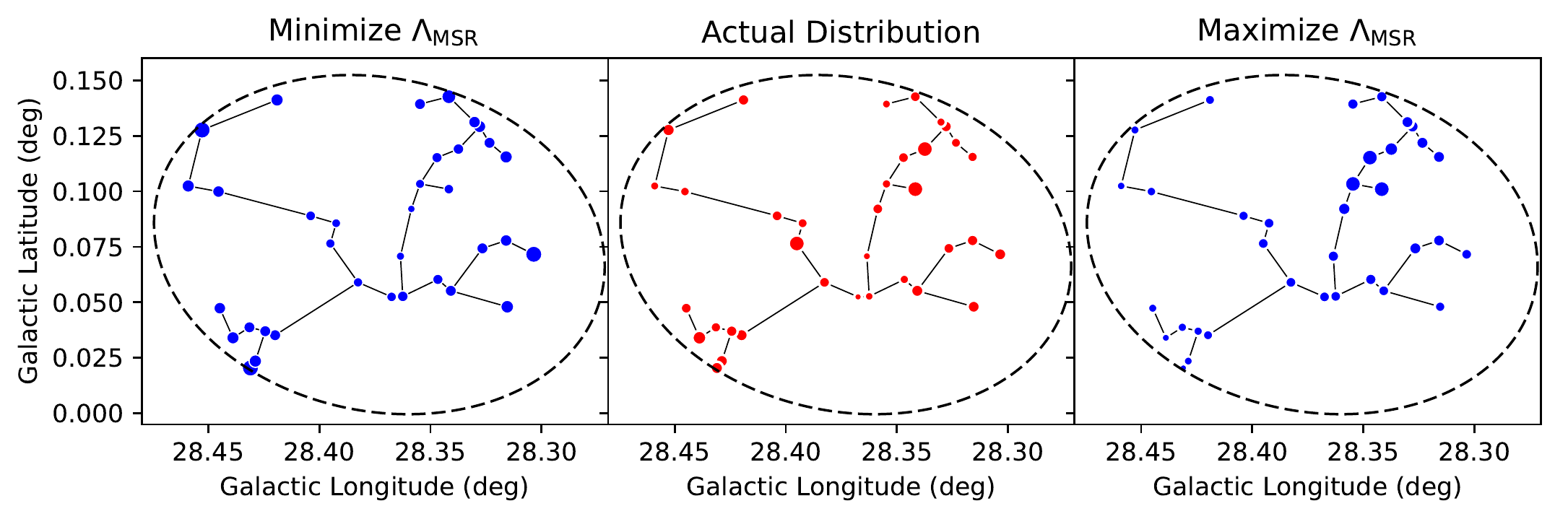}
\caption{
The three panels show the actual spatial distribution of the
protostars. The central panel shows the actual protostellar masses of
these sources (based on average of ``good'' models), with the size of
the symbols proportional to the mass. The left panel shows the
distribution of masses among the sources that minimizes the mass
segregation parameter, $\Lambda_{\rm MSR}$, i.e., an inverse mass
segregation in which the most massive stars are more separated from
each other than the typical star. The right panel shows the
distribution of masses among the sources that maximizes $\Lambda_{\rm
  MSR}$, i.e., maximum mass segregation in which the most massive
stars are least separated from each other than the typical star.
}\label{fig:msr_map}
\end{figure*}

A graphical illustration of the minimum and maximum levels of mass
segregation as measured by $\Lambda_{\rm MSR}$ is shown in
Figure~\ref{fig:msr_map}. This figure shows the protostars at their
actual locations, but with the masses free to be swapped around to
make the minimum and maximum levels of mass segregation (the area of
the symbol is proportional to the mass of the protostar). The most
mass segregated case places the most massive stars together in the
region of highest source areal density in the northwestern region of
the IRDC. The most inverse mass segregated case places the most
massive stars in a ring near the outer boundary of the IRDC. These
distributions are independent of $N$.

Our result of an apparent lack of or limited level of enhanced
clustering for more massive sources is similar to results presented by
Rom\'an-Z\'u\~{n}iga et al. (2019) for density peaks identified in
extinction maps of the Pipe nebula, but different from their results
in the Orion region, where they do find evidence for $\Lambda_{\rm
  MSR}$ rising systematically as peak mass increases. There are
important differences between our work and that of Rom\'an-Z\'u\~{n}iga et
al. (2019), including: we have considered protostellar masses, rather
than core (or peak) gas masses; our sources were identified by their
70~$\rm \mu m$ emission with the CuTEx algorithm (see \S2), while
Rom\'an-Z\'u\~{n}iga et al. used the clumpfind algorithm (Williams et
al. 1994) to find sources in their column density map; and our target
cloud, IRDC G028.37+00.07, is much more massive and of higher velocity
dispersion than the clouds studied by Rom\'an-Z\'u\~{n}iga et al. (2019). In
general, it will be important to extend these types of studies, using
uniform methods, to larger samples of clouds that probe wider ranges
of physical conditions and Galactic environments.

\section{Conclusions}\label{S:conclusions}

We have carried out a study of the protostellar population of the
massive ($\sim70,000\:M_\odot$) IRDC G028.37+00.07, identifying 35
sources based on their $70\:\mu$m emission observed by the {\it
  Herschel} telescope. We have measured the SEDs of the sources from 8
to 500~$\rm \mu m$, exploring the effects of choice of aperture size
and background subtraction. Models of protostars forming from 10 to
480~$M_\odot$ cores in dense environments, similar to that of the
IRDC, were then fit to the SEDs.

The protostars are found to have a range of isotropic luminosities
from $\sim$20 to 4,500$\:L_\odot$. The most luminous sources are
predicted to have current protostellar masses of
$m_{*}\sim10\:M_\odot$ forming from cores of mass $M_{c}\sim40$ to
$400\:M_\odot$. On the other hand, the least luminous sources are
predicted to be protostars with masses as low as $\sim 0.5\:M_\odot$
forming from cores with $M_{c}\sim10\:M_\odot$, which are at the
boundary of the protostellar model grid. We have discussed the
uncertainties in fitting the protostar models to these data, as well
as the degeneracies in the derived parameters.

We have then attempted to estimate the total protostellar population
in the IRDC, including a completeness correction based partly on a
sub-region of the IRDC that has higher angular resolution {\it ALMA}
observations sensitive to lower-mass and more embedded protostars that
are driving CO outflows. From the derived total protostellar
population we estimate a star formation efficiency per free-fall time
of $\sim3\%$ in the IRDC. Thus, if a bound cluster is to be produced,
requiring high values of total star formation efficiency of at least
30\%, then the star formation process needs to continue over about 10
current free-fall times of the cloud.

Finally, analyzing the spatial distribution of the sources, we find
that there is a high degree of substructure, similar to that found in
lower-mass protoclusters. There is also a relatively low degree of
central concentration of the protostars. The protostars, including the
most massive ones, do not appear to be especially centrally
concentrated in the protocluster as defined by the IRDC boundary,
i.e., there is no clear evidence for primordial mass segregation
in this massive IRDC.

This study is the first attempt to build a complete census of high-
and intermediate-mass star formation in a very massive early stage
protocluster. Studies of a larger number of systems are needed. The
work has illustrated the limitations of current observational
datasets, especially the relatively low angular resolution of the
infrared images from {\it Herschel} that are used to build the
SEDs. Improvements in angular resolution and sensitivity of infrared
observations, e.g., as expected from {\it JWST} and MIR observations
with {\it TMT} and {\it E-ELT}, albeit at relatively short
wavelengths, are needed to better characterize the protostellar
populations in such systems. Sub-mm observations from interferometers,
especially {\it ALMA}, may also be helpful, however, they currently
suffer from spatial filtering of flux on larger scales.

\acknowledgements J.C.T. acknowledges support from NSF grant AST~1411527 and ERC project 788829~MSTAR and VR project Fire from Ice: The Evolutionary Sequence of Massive Star Formation.

\newpage
\begin{deluxetable}{cccccccc}
\onecolumngrid
\tabletypesize{\scriptsize}
\tablecaption{Effect of Aperture Size on Protostellar Model Results\tablenotemark{a}\label{tab:aperture}}
\tablewidth{14pt}
\tablehead{
\colhead{Source} &\colhead{$\chi^{2}$} & \colhead{$M_{\rm c}$} & \colhead{$\Sigma_{\rm cl}$}  &\colhead{$m_{*}$} & \colhead{$M_{\rm env}$} & \colhead{$L_{\rm tot,iso}$} & \colhead{$L_{\rm tot,bol}$} \\
&\colhead{} & \colhead{($M_\odot$)} & \colhead{(g $\rm cm^{-2}$)} & \colhead{($M_{\odot}$)} & \colhead{($M_{\odot}$)} & \colhead{($L_{\odot}$)} & \colhead{($L_{\odot}$)} \\
}
\startdata
Cp23 
& 19.866, $12.425_{2.587}^{34.817}$  &  480, $400_{160}^{480}$ & 3.2, $3.2_{3.2}^{3.2}$ & 8.0, $8.0_{4.0}^{8.0}$ & 462, $382_{153}^{462}$  & 40000, $40000_{32000}^{40000}$ & 22000, $20000_{12000}^{22000}$ \\
$R_{\rm ap}$ = 28\arcsec
& 27.028, $15.720_{3.839}^{54.884}$ &  414, $312_{221}^{348}$ & 2.4, $3.2_{1.8}^{2.4}$ & 8.9, $8.0_{6.2}^{9.8}$ & 398, $296_{209}^{331}$ & 50000, $51000_{36000}^{55000}$ & 29000, $20000_{14000}^{31000}$ \\
\hline\noalign{\smallskip}
Cp15 
& 2.324, $0.662_{0.997}^{0.299}$ & 20, $80_{20}^{60}$ & 3.2, $0.1_{0.1}^{0.3}$ & 0.5, $1.0_{1.0}^{0.5}$ & 19, $77_{17}^{60}$ & 2800, $170_{94}^{300}$ & 860, $190_{150}^{180}$ \\
$R_{\rm ap}$ = 10\arcsec
&  3.026, $0.980_{1.364}^{0.470}$ & 126, $69_{34}^{111}$ & 0.4, $0.1_{0.1}^{0.1}$ & 1.7, $0.9_{0.7}^{0.9}$ & 123, $66_{31}^{109}$ & 1800, $150_{86}^{250}$ & 980, $180_{110}^{210}$ \\
\hline\noalign{\smallskip}
Cp03 
& 0.054, $0.248_{0.378}^{0.239}$ & 60, $10_{10}^{10}$ & 1.0, $0.3_{0.1}^{0.1}$ & 0.5, $4.0_{2.0}^{1.0}$ & 59, $1_{4}^{7}$  & 570, $49_{20}^{44}$ & 400, $670_{130}^{110}$ \\
$R_{\rm ap}$ = 9\arcsec
& 0.092, $0.399_{0.378}^{0.520}$ & 88, $15_{10}^{18}$ & 0.3, $0.1_{0.1}^{0.1}$ & 1.0, $2.9_{2.0}^{1.6}$ & 86, $3_{4}^{6}$ & 520, $45_{20}^{49}$ & 380, $340_{370}^{340}$ \\
\enddata
\tablenotetext{a}{
Within each column, the first value uses fiducial aperture size
without background subtraction, the second value is the fiducial
aperture with background subtraction, the subscripted value is the
case with aperture 30\% smaller than the fiducial, and the superscripted
value is the aperture 30\% larger than the fiducial. For each source, two lines are shown: the first line is the best fitting model; the second line is the average of ``good'' models (see text). The first column also lists the angular size of the fiducial aperture.}
\end{deluxetable}

\begin{deluxetable}{cccccccccccccc}
\onecolumngrid
\tabletypesize{\scriptsize}
\tablecaption{Parameters of the Fitted Models for all Sources}\label{tab:all}
\tablewidth{14pt}
\tablehead{
\colhead{Source} &\colhead{$R_{\rm ap}$}&\colhead{$\chi^{2}$} & \colhead{$M_{c}$} & \colhead{$\Sigma_{\rm cl}$} &\colhead{$R_{c}$}&\colhead{$m_{*}$} & \colhead{$\theta_{\rm view}$} &\colhead{$A_{V}$} & \colhead{$M_{\rm env}$} &\colhead{$\theta_{w,\rm esc}$} & \colhead{$\dot {M}_{\rm disk}$} & \colhead{$L_{\rm tot,iso}$} & \colhead{$L_{\rm tot,bol}$} \\
\colhead{} & \colhead{(arcsec)} &\colhead{} & \colhead{($M_\odot$)} & \colhead{(g $\rm cm^{-2}$)} &\colhead{pc (arcsec)} &\colhead{($M_{\odot}$)} & \colhead{(deg)} & \colhead{(mag)} & \colhead{($M_{\odot}$)} & \colhead{(deg)} &\colhead{($M_{\odot}$/yr)} & \colhead{($L_{\odot}$)} & \colhead{($L_{\odot}$)} \\
}
\startdata
Cp01
& 15 & 0.439 & 10 & 3.2 & 0.013 ( 0.54 ) & 4.0 & 64.85 & 16.2 & 2 & 56 & 1.9$\rm \times 10^{-4}$ & 2.4$\rm \times 10^{2}$ & 1.9$\rm \times 10^{3}$ \\
& 10 & 0.908 & 23 & 0.3 & 0.063 ( 2.59 ) & $2.64 ^{ 7.45 }_{ 0.94 }$ & 59.69 & 36.8 & 10 & 37 & 4.0$\rm \times 10^{-5}$ & 3.8$\rm \times 10^{2}$ & 1.1$\rm \times 10^{3}$ \\
\hline\noalign{\smallskip}
Cp02
& 6 & 0.955 & 10 & 0.3 & 0.041 ( 1.71 ) & 4.0 & 88.57 & 100.0 & 1 & 68 & 2.4$\rm \times 10^{-5}$ & 2.9$\rm \times 10^{1}$ & 6.7$\rm \times 10^{2}$ \\
& 4 & 1.235 & 19 & 0.2 & 0.075 ( 3.11 ) & $5.83 ^{ 12.50 }_{ 2.72 }$ & 88.57 & 100.0 & 1 & 70 & 1.6$\rm \times 10^{-5}$ & 3.9$\rm \times 10^{1}$ & 4.7$\rm \times 10^{2}$ \\
\hline\noalign{\smallskip}
Cp03
& 9 & 0.248 & 10 & 0.3 & 0.041 ( 1.71 ) & 4.0 & 77.00 & 100.0 & 1 & 68 & 2.4$\rm \times 10^{-5}$ & 4.9$\rm \times 10^{1}$ & 6.7$\rm \times 10^{2}$ \\
& 6 & 0.399 & 15 & 0.1 & 0.075 ( 3.08 ) & $2.88 ^{ 9.48 }_{ 0.88 }$ & 85.20 & 84.2 & 3 & 55 & 1.3$\rm \times 10^{-5}$ & 4.5$\rm \times 10^{1}$ & 3.4$\rm \times 10^{2}$ \\
\hline\noalign{\smallskip}
Cp04
& 12 & 0.235 & 20 & 0.3 & 0.059 ( 2.42 ) & 4.0 & 82.82 & 0.0 & 11 & 38 & 5.4$\rm \times 10^{-5}$ & 3.1$\rm \times 10^{2}$ & 1.1$\rm \times 10^{3}$ \\
& 10 & 0.569 & 34 & 0.2 & 0.102 ( 4.22 ) & $2.46 ^{ 3.84 }_{ 1.58 }$ & 64.63 & 23.7 & 26 & 24 & 3.5$\rm \times 10^{-5}$ & 3.3$\rm \times 10^{2}$ & 5.6$\rm \times 10^{2}$ \\
\hline\noalign{\smallskip}
Cp05
& 10 & 0.109 & 10 & 0.3 & 0.041 ( 1.71 ) & 2.0 & 88.57 & 84.8 & 5 & 43 & 3.0$\rm \times 10^{-5}$ & 5.8$\rm \times 10^{1}$ & 2.8$\rm \times 10^{2}$ \\
& 10 & 0.311 & 22 & 0.1 & 0.103 ( 4.27 ) & $0.76 ^{ 1.32 }_{ 0.44 }$ & 82.85 & 89.6 & 19 & 19 & 1.3$\rm \times 10^{-5}$ & 7.0$\rm \times 10^{1}$ & 1.1$\rm \times 10^{2}$ \\
\hline\noalign{\smallskip}
Cp06
& 9 & 0.014 & 10 & 0.3 & 0.041 ( 1.71 ) & 4.0 & 82.82 & 100.0 & 1 & 68 & 2.4$\rm \times 10^{-5}$ & 3.4$\rm \times 10^{1}$ & 6.7$\rm \times 10^{2}$ \\
& 10 & 0.087 & 16 & 0.1 & 0.079 ( 3.28 ) & $1.89 ^{ 5.93 }_{ 0.60 }$ & 87.71 & 91.3 & 5 & 43 & 1.4$\rm \times 10^{-5}$ & 5.2$\rm \times 10^{1}$ & 3.7$\rm \times 10^{2}$ \\
\hline\noalign{\smallskip}
Cp07
& 8 & 0.159 & 10 & 0.3 & 0.041 ( 1.71 ) & 2.0 & 77.00 & 100.0 & 5 & 43 & 3.0$\rm \times 10^{-5}$ & 6.0$\rm \times 10^{1}$ & 2.8$\rm \times 10^{2}$ \\
& 10 & 0.307 & 20 & 0.1 & 0.093 ( 3.85 ) & $1.43 ^{ 4.66 }_{ 0.44 }$ & 72.51 & 80.9 & 9 & 34 & 1.3$\rm \times 10^{-5}$ & 1.9$\rm \times 10^{2}$ & 3.0$\rm \times 10^{2}$ \\
\hline\noalign{\smallskip}
Cp08
& 10 & 0.445 & 10 & 0.3 & 0.041 ( 1.71 ) & 0.5 & 85.70 & 100.0 & 9 & 18 & 1.9$\rm \times 10^{-5}$ & 1.2$\rm \times 10^{2}$ & 1.9$\rm \times 10^{2}$ \\
& 10 & 0.659 & 44 & 0.1 & 0.138 ( 5.68 ) & $0.76 ^{ 1.20 }_{ 0.48 }$ & 69.79 & 67.9 & 40 & 12 & 1.7$\rm \times 10^{-5}$ & 1.1$\rm \times 10^{2}$ & 1.5$\rm \times 10^{2}$ \\
\hline\noalign{\smallskip}
Cp09
& 20 & 7.383 & 480 & 0.1 & 0.510 ( 21.03 ) & 8.0 & 28.96 & 10.1 & 463 & 9 & 8.5$\rm \times 10^{-5}$ & 9.3$\rm \times 10^{3}$ & 9.7$\rm \times 10^{3}$ \\
& 6 & 10.082 & 376 & 0.1 & 0.451 ( 18.62 ) & $10.48 ^{ 12.69 }_{ 8.66 }$ & 28.76 & 66.8 & 351 & 13 & 9.1$\rm \times 10^{-5}$ & 1.5$\rm \times 10^{4}$ & 1.7$\rm \times 10^{4}$ \\
\hline\noalign{\smallskip}
Cp10
& 6 & 1.040 & 10 & 0.1 & 0.074 ( 3.04 ) & 2.0 & 88.57 & 100.0 & 4 & 50 & 1.1$\rm \times 10^{-5}$ & 2.0$\rm \times 10^{1}$ & 1.3$\rm \times 10^{2}$ \\
& 2 & 1.189 & 10 & 0.2 & 0.055 ( 2.28 ) & $2.83 ^{ 4.00 }_{ 2.00 }$ & 88.57 & 100.0 & 2 & 59 & 1.7$\rm \times 10^{-5}$ & 2.4$\rm \times 10^{1}$ & 3.7$\rm \times 10^{2}$ \\
\hline\noalign{\smallskip}
Cp11
& 8 & 0.868 & 10 & 1.0 & 0.023 ( 0.96 ) & 2.0 & 43.53 & 16.2 & 5 & 39 & 7.5$\rm \times 10^{-5}$ & 2.6$\rm \times 10^{2}$ & 7.6$\rm \times 10^{2}$ \\
& 7 & 1.291 & 16 & 0.4 & 0.048 ( 1.97 ) & $2.69 ^{ 7.60 }_{ 0.95 }$ & 58.95 & 45.3 & 6 & 43 & 4.0$\rm \times 10^{-5}$ & 3.9$\rm \times 10^{2}$ & 7.9$\rm \times 10^{2}$ \\
\hline\noalign{\smallskip}
Cp12
& 19 & 0.265 & 60 & 1.0 & 0.057 ( 2.35 ) & 24.0 & 88.57 & 42.4 & 5 & 71 & 1.9$\rm \times 10^{-4}$ & 2.1$\rm \times 10^{3}$ & 9.3$\rm \times 10^{4}$ \\
& 10 & 0.708 & 38 & 0.4 & 0.076 ( 3.13 ) & $9.80 ^{ 20.40 }_{ 4.71 }$ & 68.88 & 56.9 & 10 & 54 & 7.8$\rm \times 10^{-5}$ & 2.5$\rm \times 10^{3}$ & 1.4$\rm \times 10^{4}$ \\
\hline\noalign{\smallskip}
Cp13
& 12 & 10.302 & 240 & 0.1 & 0.360 ( 14.87 ) & 1.0 & 12.84 & 60.6 & 240 & 4 & 2.6$\rm \times 10^{-5}$ & 3.2$\rm \times 10^{2}$ & 2.4$\rm \times 10^{2}$ \\
& 10 & 11.740 & 173 & 0.1 & 0.306 ( 12.63 ) & $1.52 ^{ 2.13 }_{ 1.08 }$ & 37.76 & 79.7 & 169 & 7 & 2.9$\rm \times 10^{-5}$ & 3.4$\rm \times 10^{2}$ & 3.0$\rm \times 10^{2}$ \\
\hline\noalign{\smallskip}
Cp14
& 10 & 0.013 & 10 & 0.3 & 0.041 ( 1.71 ) & 4.0 & 88.57 & 78.8 & 1 & 68 & 2.4$\rm \times 10^{-5}$ & 2.9$\rm \times 10^{1}$ & 6.7$\rm \times 10^{2}$ \\
& 10 & 0.136 & 16 & 0.1 & 0.079 ( 3.28 ) & $1.89 ^{ 5.93 }_{ 0.60 }$ & 88.57 & 89.1 & 5 & 43 & 1.4$\rm \times 10^{-5}$ & 5.1$\rm \times 10^{1}$ & 3.7$\rm \times 10^{2}$ \\
\hline\noalign{\smallskip}
Cp15
& 10 & 0.662 & 80 & 0.1 & 0.208 ( 8.59 ) & 1.0 & 88.57 & 100.0 & 77 & 8 & 1.9$\rm \times 10^{-5}$ & 1.7$\rm \times 10^{2}$ & 1.9$\rm \times 10^{2}$ \\
& 10 & 0.980 & 69 & 0.1 & 0.183 ( 7.53 ) & $0.87 ^{ 1.32 }_{ 0.57 }$ & 62.06 & 81.2 & 66 & 10 & 1.9$\rm \times 10^{-5}$ & 1.5$\rm \times 10^{2}$ & 1.8$\rm \times 10^{2}$ \\
\hline\noalign{\smallskip}
Cp16
& 6 & 8.522 & 10 & 1.0 & 0.023 ( 0.96 ) & 4.0 & 88.57 & 100.0 & 1 & 59 & 7.7$\rm \times 10^{-5}$ & 1.1$\rm \times 10^{2}$ & 1.1$\rm \times 10^{3}$ \\
& 9 & 12.995 & 14 & 0.3 & 0.051 ( 2.12 ) & $1.59 ^{ 3.53 }_{ 0.71 }$ & 83.23 & 62.4 & 7 & 35 & 2.8$\rm \times 10^{-5}$ & 1.1$\rm \times 10^{2}$ & 2.9$\rm \times 10^{2}$ \\
\hline\noalign{\smallskip}
Cp17
& 8 & 0.068 & 10 & 0.1 & 0.074 ( 3.04 ) & 1.0 & 88.57 & 100.0 & 7 & 31 & 1.0$\rm \times 10^{-5}$ & 4.4$\rm \times 10^{1}$ & 1.1$\rm \times 10^{2}$ \\
& 9 & 0.608 & 16 & 0.1 & 0.077 ( 3.18 ) & $1.88 ^{ 6.27 }_{ 0.56 }$ & 85.99 & 93.6 & 4 & 44 & 1.3$\rm \times 10^{-5}$ & 5.1$\rm \times 10^{1}$ & 3.7$\rm \times 10^{2}$ \\
\hline\noalign{\smallskip}
Cp18
& 10 & 1.308 & 60 & 0.3 & 0.101 ( 4.18 ) & 0.5 & 12.84 & 84.8 & 60 & 5 & 3.0$\rm \times 10^{-5}$ & 3.0$\rm \times 10^{2}$ & 1.8$\rm \times 10^{2}$ \\
& 9 & 1.617 & 115 & 0.1 & 0.206 ( 8.50 ) & $0.79 ^{ 1.10 }_{ 0.57 }$ & 32.13 & 96.1 & 113 & 5 & 2.5$\rm \times 10^{-5}$ & 2.5$\rm \times 10^{2}$ & 2.1$\rm \times 10^{2}$ \\
\hline\noalign{\smallskip}
Cp19
& 8 & 0.764 & 120 & 0.1 & 0.255 ( 10.52 ) & 0.5 & 85.70 & 100.0 & 118 & 4 & 1.5$\rm \times 10^{-5}$ & 8.4$\rm \times 10^{1}$ & 8.8$\rm \times 10^{1}$ \\
& 10 & 1.029 & 53 & 0.1 & 0.160 ( 6.58 ) & $0.57 ^{ 0.76 }_{ 0.44 }$ & 79.12 & 100.0 & 50 & 9 & 1.4$\rm \times 10^{-5}$ & 8.9$\rm \times 10^{1}$ & 1.1$\rm \times 10^{2}$ \\
\hline\noalign{\smallskip}
Cp20
& 10 & 0.118 & 120 & 0.1 & 0.255 ( 10.52 ) & 0.5 & 85.70 & 100.0 & 118 & 4 & 1.5$\rm \times 10^{-5}$ & 8.4$\rm \times 10^{1}$ & 8.8$\rm \times 10^{1}$ \\
& 10 & 0.176 & 51 & 0.1 & 0.156 ( 6.45 ) & $0.54 ^{ 0.66 }_{ 0.44 }$ & 87.71 & 100.0 & 48 & 9 & 1.4$\rm \times 10^{-5}$ & 8.3$\rm \times 10^{1}$ & 1.0$\rm \times 10^{2}$ \\
\hline\noalign{\smallskip}
Cp21
& 8 & 1.165 & 30 & 0.3 & 0.072 ( 2.96 ) & 12.0 & 88.57 & 41.4 & 1 & 81 & 2.2$\rm \times 10^{-5}$ & 7.0$\rm \times 10^{1}$ & 1.2$\rm \times 10^{4}$ \\
& 5 & 1.378 & 16 & 0.2 & 0.075 ( 3.09 ) & $4.10 ^{ 10.92 }_{ 1.54 }$ & 86.27 & 34.3 & 2 & 63 & 1.4$\rm \times 10^{-5}$ & 4.1$\rm \times 10^{1}$ & 4.7$\rm \times 10^{2}$ \\
\hline\noalign{\smallskip}
Cp22
& 8 & 0.009 & 80 & 0.1 & 0.208 ( 8.59 ) & 1.0 & 12.84 & 92.9 & 77 & 8 & 1.9$\rm \times 10^{-5}$ & 4.6$\rm \times 10^{2}$ & 1.9$\rm \times 10^{2}$ \\
& 10 & 0.022 & 65 & 0.1 & 0.188 ( 7.75 ) & $1.23 ^{ 1.69 }_{ 0.90 }$ & 21.73 & 66.3 & 61 & 12 & 2.0$\rm \times 10^{-5}$ & 2.6$\rm \times 10^{2}$ & 2.1$\rm \times 10^{2}$ \\
\hline\noalign{\smallskip}
Cp23
& 28 & 12.425 & 400 & 3.2 & 0.083 ( 3.42 ) & 8.0 & 12.84 & 36.4 & 382 & 7 & 1.1$\rm \times 10^{-3}$ & 4.2$\rm \times 10^{4}$ & 2.0$\rm \times 10^{4}$ \\
& 5 & 15.720 & 312 & 3.2 & 0.073 ( 3.02 ) & $8.00 ^{ 8.00 }_{ 8.00 }$ & 12.84 & 59.4 & 296 & 9 & 1.0$\rm \times 10^{-3}$ & 5.1$\rm \times 10^{4}$ & 2.0$\rm \times 10^{4}$ \\
\hline\noalign{\smallskip}
Cp24
& 8 & 1.066 & 10 & 0.1 & 0.074 ( 3.04 ) & 0.5 & 54.90 & 100.0 & 9 & 20 & 7.8e-06 & 5.0$\rm \times 10^{1}$ & 7.5$\rm \times 10^{1}$ \\
& 10 & 1.289 & 22 & 0.1 & 0.098 ( 4.03 ) & $1.04 ^{ 3.04 }_{ 0.36 }$ & 72.82 & 100.0 & 11 & 29 & 1.3$\rm \times 10^{-5}$ & 1.2$\rm \times 10^{2}$ & 2.0$\rm \times 10^{2}$ \\
\hline\noalign{\smallskip}
Cp25
& 14 & 0.516 & 40 & 1.0 & 0.047 ( 1.92 ) & 1.0 & 12.84 & 38.4 & 39 & 10 & 9.1$\rm \times 10^{-5}$ & 3.6$\rm \times 10^{3}$ & 1.0$\rm \times 10^{3}$ \\
& 10 & 0.970 & 54 & 0.7 & 0.064 ( 2.64 ) & $1.74 ^{ 3.43 }_{ 0.88 }$ & 17.56 & 34.6 & 49 & 13 & 9.8$\rm \times 10^{-5}$ & 2.5$\rm \times 10^{3}$ & 1.2$\rm \times 10^{3}$ \\
\hline\noalign{\smallskip}
Cp26
& 10 & 0.282 & 10 & 0.3 & 0.041 ( 1.71 ) & 1.0 & 88.57 & 25.3 & 8 & 28 & 2.5$\rm \times 10^{-5}$ & 1.1$\rm \times 10^{2}$ & 2.6$\rm \times 10^{2}$ \\
& 10 & 0.660 & 17 & 0.2 & 0.064 ( 2.65 ) & $1.41 ^{ 2.88 }_{ 0.70 }$ & 74.50 & 78.1 & 11 & 28 & 2.5$\rm \times 10^{-5}$ & 1.3$\rm \times 10^{2}$ & 3.0$\rm \times 10^{2}$ \\
\hline\noalign{\smallskip}
Cp27
& 10 & 0.102 & 10 & 0.3 & 0.041 ( 1.71 ) & 4.0 & 85.70 & 48.5 & 1 & 68 & 2.4$\rm \times 10^{-5}$ & 3.0$\rm \times 10^{1}$ & 6.7$\rm \times 10^{2}$ \\
& 10 & 0.165 & 14 & 0.3 & 0.054 ( 2.25 ) & $2.86 ^{ 7.38 }_{ 1.11 }$ & 87.71 & 65.2 & 3 & 52 & 2.3$\rm \times 10^{-5}$ & 5.8$\rm \times 10^{1}$ & 6.4$\rm \times 10^{2}$ \\
\hline\noalign{\smallskip}
Cp28
& 8 & 0.108 & 10 & 0.3 & 0.041 ( 1.71 ) & 4.0 & 88.57 & 100.0 & 1 & 68 & 2.4$\rm \times 10^{-5}$ & 2.9$\rm \times 10^{1}$ & 6.7$\rm \times 10^{2}$ \\
& 10 & 0.663 & 14 & 0.3 & 0.054 ( 2.25 ) & $2.49 ^{ 7.35 }_{ 0.84 }$ & 87.71 & 99.1 & 3 & 50 & 2.1$\rm \times 10^{-5}$ & 5.7$\rm \times 10^{1}$ & 5.9$\rm \times 10^{2}$ \\
\hline\noalign{\smallskip}
Cp29
& 8 & 0.119 & 10 & 0.3 & 0.041 ( 1.71 ) & 4.0 & 79.92 & 100.0 & 1 & 68 & 2.4$\rm \times 10^{-5}$ & 4.0$\rm \times 10^{1}$ & 6.7$\rm \times 10^{2}$ \\
& 10 & 0.282 & 14 & 0.3 & 0.054 ( 2.25 ) & $2.86 ^{ 7.38 }_{ 1.11 }$ & 87.13 & 100.0 & 3 & 52 & 2.3$\rm \times 10^{-5}$ & 6.0$\rm \times 10^{1}$ & 6.4$\rm \times 10^{2}$ \\
\hline\noalign{\smallskip}
Cp30
& 11 & 0.016 & 10 & 3.2 & 0.013 ( 0.54 ) & 4.0 & 82.82 & 9.1 & 2 & 56 & 1.9$\rm \times 10^{-4}$ & 1.6$\rm \times 10^{2}$ & 1.9$\rm \times 10^{3}$ \\
& 10 & 0.044 & 11 & 0.3 & 0.044 ( 1.83 ) & $1.62 ^{ 3.48 }_{ 0.76 }$ & 55.30 & 27.7 & 6 & 37 & 2.9$\rm \times 10^{-5}$ & 1.2$\rm \times 10^{2}$ & 3.5$\rm \times 10^{2}$ \\
\hline\noalign{\smallskip}
Cp31
& 20 & 0.921 & 320 & 0.1 & 0.416 ( 17.17 ) & 8.0 & 88.57 & 72.7 & 307 & 11 & 7.7$\rm \times 10^{-5}$ & 7.5$\rm \times 10^{3}$ & 8.8$\rm \times 10^{3}$ \\
& 8 & 1.234 & 139 & 0.4 & 0.134 ( 5.52 ) & $4.76 ^{ 8.45 }_{ 2.68 }$ & 37.62 & 74.7 & 129 & 12 & 1.4$\rm \times 10^{-4}$ & 1.2$\rm \times 10^{4}$ & 6.8$\rm \times 10^{3}$ \\
\hline\noalign{\smallskip}
Cp32
& 11 & 0.409 & 10 & 0.1 & 0.074 ( 3.04 ) & 0.5 & 88.57 & 100.0 & 9 & 20 & 7.8e-06 & 4.6$\rm \times 10^{1}$ & 7.5$\rm \times 10^{1}$ \\
& 10 & 0.643 & 18 & 0.1 & 0.094 ( 3.86 ) & $1.20 ^{ 3.12 }_{ 0.46 }$ & 80.42 & 100.0 & 10 & 31 & 1.2$\rm \times 10^{-5}$ & 8.2$\rm \times 10^{1}$ & 1.9$\rm \times 10^{2}$ \\
\hline\noalign{\smallskip}
Cp33
& 9 & 0.745 & 10 & 0.1 & 0.074 ( 3.04 ) & 0.5 & 88.57 & 100.0 & 9 & 20 & 7.8e-06 & 4.6$\rm \times 10^{1}$ & 7.5$\rm \times 10^{1}$ \\
& 5 & 1.068 & 10 & 0.2 & 0.058 ( 2.41 ) & $1.52 ^{ 3.07 }_{ 0.75 }$ & 85.68 & 92.5 & 4 & 43 & 1.5$\rm \times 10^{-5}$ & 3.8$\rm \times 10^{1}$ & 2.1$\rm \times 10^{2}$ \\
\hline\noalign{\smallskip}
Cp34
& 10 & 6.589 & 10 & 1.0 & 0.023 ( 0.96 ) & 4.0 & 12.84 & 0.0 & 1 & 59 & 7.7$\rm \times 10^{-5}$ & 3.4$\rm \times 10^{3}$ & 1.1$\rm \times 10^{3}$ \\
& 5 & 9.966 & 23 & 0.4 & 0.056 ( 2.29 ) & $3.78 ^{ 7.28 }_{ 1.96 }$ & 21.25 & 0.0 & 5 & 46 & 4.4$\rm \times 10^{-5}$ & 3.2$\rm \times 10^{3}$ & 1.7$\rm \times 10^{3}$ \\
\hline\noalign{\smallskip}
Cp35
& 8 & 1.991 & 10 & 0.3 & 0.041 ( 1.71 ) & 1.0 & 88.57 & 60.6 & 8 & 28 & 2.5$\rm \times 10^{-5}$ & 1.1$\rm \times 10^{2}$ & 2.6$\rm \times 10^{2}$ \\
& 2 & 2.592 & 10 & 0.3 & 0.041 ( 1.71 ) & $0.71 ^{ 1.00 }_{ 0.50 }$ & 58.76 & 80.3 & 8 & 23 & 2.1$\rm \times 10^{-5}$ & 1.2$\rm \times 10^{2}$ & 2.4$\rm \times 10^{2}$ \\
\enddata
\tablenotetext{a}{
For each source, two lines are shown: the first line is the best fitting model; the second line is the average of ``good'' models (see text). For the current protostellar masses of the good models, we also indicate with super- and subscripts the range of masses set by $\pm1\sigma$, where $\sigma$ is the standard deviation in log$_{10}\:m_*$.}
\end{deluxetable}

\end{document}